\documentclass[12pt,letterpaper]{article}

\usepackage[includeheadfoot,
            marginratio={1:1,2:3}, 
            width=412pt, 
            height=688pt,]{geometry}

\usepackage{amsmath}
\usepackage{amsfonts}
\usepackage{amssymb}
\usepackage{graphicx}
\usepackage{cite}
\usepackage{ulem}
\usepackage{dsfont}
\usepackage{longtable}
\usepackage{afterpage}
\usepackage{multirow}
\usepackage{url}

\newcommand{\eq}[1]{\begin{equation}
                     \begin{split} #1 \end{split}
                     \end{equation}}

\newcommand{\ov}{\overline}
\newcommand{\op}{\hspace{1pt}}

\allowdisplaybreaks[2]
\numberwithin{equation}{section}

\begin{document}

\normalem
\vspace*{-1.5cm}
\begin{flushright}
  {\small
  MPP-2016-323 \\
  LMU-ASC 55/16 
  }
\end{flushright}

\vspace{1.5cm}

\begin{center}
  {\LARGE
    The Asymmetric CFT Landscape  in $D=4,6,8$\\[0.2cm]
    with Extended  Supersymmetry
}
\vspace{0.4cm}

\end{center}

\vspace{0.35cm}
\begin{center}
  Ralph Blumenhagen$^1$, Michael Fuchs$^1$, Erik Plauschinn$^2$
\end{center}

\vspace{0.1cm}
\begin{center} 
\emph{
$^{1}$ Max-Planck-Institut f\"ur Physik (Werner-Heisenberg-Institut), \\ 
F\"ohringer Ring 6,  80805 M\"unchen, Germany 
} \\[1em] 

\emph{
$^{2}$ Arnold Sommerfeld Center for Theoretical Physics,\\ 
LMU, Theresienstr.~37, 80333 M\"unchen, Germany
}
\end{center} 

\vspace{1cm}


\begin{abstract}
\noindent
We study  asymmetric simple-current extensions of Gepner models in dimensions 
$D=4,6,8$ with at least eight supercharges in the right-moving sector. 
The models obtained in an extensive stochastic computer search 
belong to a small number of different classes. 
These classes can be categorized as dimensional
reductions, asymmetric orbifolds with $(-1)^{F_L}$, extra gauge
enhancement and as coming from the super Higgs-effect. 
Models in the latter class are particularly interesting, as they may correspond to
non-geometric flux compactifications. 
\end{abstract}


\clearpage
\tableofcontents


\section{Introduction}

Much of the work on string compactifications is concerned with  
the description and understanding of geometric backgrounds.
Examples are type II and heterotic models compactified on Calabi-Yau
manifolds, left-right symmetric orbifolds and orientifolds thereof.
However,  from the early days of string theory on it has been clear that 
also fully consistent {\em left-right asymmetric} conformal field theories (ACFTs)
exist. The most prominent 
examples are asymmetric orbifolds \cite{Narain:1986qm} 
and free-fermion constructions \cite{Antoniadis:1985az,Antoniadis:1986rn,Ferrara:1989nm},
and for a discussion of D-branes in asymmetric backgrounds see \cite{Brunner:1999fj,Gutperle:2000bf,Bianchi:2008cj,Bianchi:2009mu}.
More recently it has become clear that at least some of these
asymmetric constructions correspond to
NS-NS flux compactifications, in particular, they correspond to Minkowski minima of
gauged supergravity (GSUGRA) theories with spontaneously-,  partially-broken
supersymmetry \cite{Dabholkar:2002sy,Hellerman:2002ax,Flournoy:2004vn,Flournoy:2005xe,Hellerman:2006tx,Condeescu:2012sp,Condeescu:2013yma,Blumenhagen:2016axv}. The general GSUGRA theory  involves
non-geometric fluxes that naturally appear in 
double field theory, which is a proposed field theory that features manifest
$O(D,D)$ invariance (for reviews see
\cite{Aldazabal:2013sca,Berman:2013eva,Hohm:2013bwa}). 
However, it is fair to say that  the constraints on  such non-geometric
fluxes are not yet fully understood, i.e. it is not clear which minima of GSUGRA belong to the 
string landscape and which to the swampland.

This work can be considered as the continuation of a series of papers
that started with the construction of Gepner models \cite{Gepner:1987qi,Gepner:1987vz,Fuchs:1989pt,Fuchs:1989yv}, their extension by the simple current technique
\cite{Schellekens:1989am,Schellekens:1989dq} for constructing ACFTs
\cite{Schellekens:1989wx,GatoRivera:2010gv,GatoRivera:2010xn,Israel:2013wwa,Israel:2015efa}, and
the recent attempt \cite{Blumenhagen:2016axv} to relate some of these four-dimensional
ACFTs to minima of ${\cal N}=2$ GSUGRA partially-broken to 
\mbox{${\cal N}=1$}.\footnote{Similar asymmetric simple current extensions in the
context of the heterotic string were 
discussed in
\cite{Blumenhagen:1995tt,Blumenhagen:1995ew,Blumenhagen:1996vu}.}
For instance, it was proposed that the $\mathbf k=(3^5)$ Gepner model,
extended by a certain asymmetric simple current, corresponds to a
Minkowski minimum of a non-geometric flux compactification on the
complete intersection Calabi-Yau threefold $\mathbb P_{1,1,1,1,2,2}[5,3]$.
The latter identification was impeded by the existence of a superpotential in four-dimensional ${\cal N}=1$ 
theories, due to which an arbitrary  number of chiral fields can
become massive. For this reason, it is desirable to investigate the
analogous ACFT-GSUGRA correspondence in a simpler setting, where 
supersymmetry is more strongly protecting the generation of mass terms. 

Therefore, in this paper we consider ACFTs in the framework of the type IIB
superstring theory in dimensions $D=4,6,8$, constructed by extensions via simple 
currents with at least  {\em eight} supercharges arising from the
right-moving sector. Thus, the models have at least ${\cal N}=2$ in 4D
or ${\cal N}=1$ in 6D and 8D.  Geometric compactifications
belonging to these classes are on $\mathbb T^2$, $K3$ and $K3\times \mathbb T^2$, respectively. 
However, we also found models with e.g. ${\cal N}=5$ or ${\cal N}=3$ supersymmetry in 4D that
clearly can  only appear in asymmetric constructions.
Since we are working with models with extended supersymmetry, we
expect that the identification of these abstractly defined ACFTs  is
simpler than 
in our earlier work \cite{Blumenhagen:2016axv}. This indeed turns out to be the case in an unprecedented clarity.

Through an extensive stochastic computer search, we have constructed 
hundreds of millions of different ACFTs and found
that they can be understood via  four different mechanisms.
In essence,  our classification exploits the fact that in theories with eight
supercharges, no generic scalar potential exists and therefore
modes can become massive only via a Higgs mechanism. 
The four mechanisms are characterized as follows:
\begin{itemize}

\item First, some of the lower-dimensional models are simply dimensional reductions
of higher-dimensional models. 

\item Second, as it often
happens for Gepner models, there can be special gauge enhancements
that can be Higgsed. 

\item Third, even though R-R fluxes are not expected to
be visible in a CFT, it turns  out that the asymmetric
operation $(-1)^{F_L}$ can be realized via simple currents. Here
$F_L$ is the left-moving space-time fermion number, that is even for 
the left-moving NS-sector and odd for the R-sector.
In all dimensions  we found a class of models that can be identified with asymmetric
$(-1)^{F_L}$ shift orbifolds with non-abelian gauge symmetries. Such
models were already considered mostly in 4D in \cite{Dixon:1987yp,Bluhm:1988mh,Anastasopoulos:2009kj}.

\item Finally and most importantly, 
the most extensive  series of models we found can be understood via the
super Higgs mechanism \cite{Deser:1977uq,Cremmer:1978iv,Cremmer:1978hn,Andrianopoli:2002rm}. 
This mechanism determines how
a supergravity theory with ${\cal N}'$ supersymmetries can be broken to a theory with ${\mathcal N}< \mathcal N'$ supersymmetries,
consistent with the multiplet structure of both theories.
This puts a number of constraints on the massless spectra
of ${\cal N}$-supergravity models, which  can be considered as necessary
conditions for the gauged ${\cal N}'$-supergravity theory to admit
Minkowski minima with ${\cal N}$-supersymmetry. For
extended supersymmetries in 4D this was discussed in some detail in
\cite{Andrianopoli:2002rm}. Some of these models can also be realized
as asymmetric orbifolds that involve  left-right asymmetric discrete 
symmetries but no $(-1)^{F_L}$ factor.

\end{itemize}
Exploiting these four mechanisms, we are able to provide 
a classification of all  ACFT models found in our computer search.
The emerging picture is quite compelling, but due to the stochastic nature of our search we cannot 
claim  completeness of the ACFT landscape.
In particular, there can exist islands \cite{Dabholkar:1998kv} which cannot be reached
via the simple-current technique. Moreover, as will be discussed, in cases
where a super Higgs mechanism can work, i.e. where Minkowski type flux
vacua can exist in principle, fairly large classes of ACFTs do
appear. We are not yet at the stage where we can provide a one-to-one
correspondence between  asymmetric CFT data and concrete gaugings or
fluxes, but the results look encouraging.

\bigskip
This paper is organized as follows: In section~\ref{sec_gepner} we briefly introduce
Gepner models and their simple current extensions. 
Section~\ref{sec_landscape} is the main 
section and contains the presentation and classification of the
ACFTs we found in our stochastic search. For each class we only
present a typical representative example. More details on these
models can be found at the URL-link \cite{examples}. Section~\ref{sec_concl} contains
our conclusions, and the appendix contains an overview of  
multiplets in extended supersymmetries in $D=4,6,8$ dimensions.


\section{The ACFT construction}
\label{sec_gepner}
In this  section, we briefly review the asymmetric conformal field theory construction employed in this paper. This is meant to explain our procedure to find models, the notation and the adjustments one has to make when working in different dimensions. For a more detailed explanation we would like to refer
the reader for instance to \cite{Blumenhagen:2009zz} and to our previous paper \cite{Blumenhagen:2016axv}.

\subsection{Simple current extension}
In many rational conformal field theories there exist primary fields $J_a$ called simple currents \cite{Schellekens:1989am,Schellekens:1989dq}, whose fusion with any primary $\phi_i$ gives exactly one primary field, i.e.  
\eq{ J_a \times \phi_i = \phi_{J(i)} .
} 
By associativity it is easy to see that the fusion of two simple currents is a simple current as well. Furthermore, having finitely many primaries it is clear that $J_a^{ {\cal N}_a} = 1$ for a certain length ${\cal N}_a$. As a consequence the simple currents group the primaries into orbits $\{ \phi_i, J_a \times \phi_i, J_a^2 \times \phi_i, \dots , J_a^{ {\cal N}_a - 1} \times \phi_i \}$. The above fusion rules result in the operator product expansion
\eq{ 
J_a(z) \,  \phi_i(w) = (z - w )^{ - Q^{(a)}_i} \phi_{J(i)} (w)+\ldots \,,
}
where $Q_i^{(a)}$ is called the monodromy charge. Using $J_a^{ {\cal N}_a} = 1$ in this OPE one easily finds $ Q^{(a)}_i = {t^i_a\over {\cal N}_a}\ {\rm mod}\ 1$ with an integer $t^i_a$. Furthermore, applying the simple current to an arbitrary OPE shows that the monodromy charge is a conserved quantity. If a so called monodromy parameter is even (for details consult e.g. \cite{Schellekens:1989am,Schellekens:1989dq,Blumenhagen:2009zz,Blumenhagen:2016axv}), a simple current implies the existence of the following off-diagonal modular invariant partition function
\eq{
Z_a ( \tau, \bar{\tau})  =  \vec\chi^{\,T}(\tau)\, M(J_a)\,\vec\chi(\ov\tau)=
\sum_{k,l} \chi_k (\tau)\,\, (M_a)_{kl}\,\,
\chi_l (\bar{\tau})\,,
} 
where 
\eq{
(M_a)_{kl} = \sum_{p = 1}^{{\cal N}_a} \delta( \phi_k, J_a^p \times \phi_l) \, \, \, 
\delta^{(1)} \big(\hat{Q}^{(a)}(\phi_k) + \hat{Q}^{(a)} (\phi_l) \big) \, 
}
with
\eq{
\hat{Q}^{(a)} (\phi_i) = {t_a^i \over 2 \op {\cal N}_a} \ {\rm mod}\ 1 \,. 
}
Notice that the simple current takes the primaries of the left side and couples them to their whole orbit (if the monodromy charges fit). Of course the combination of several modular matrices like $Z_{a_1, a_2} = {1\over N}\sum_{k,l,m} \chi_l \, (M_{a_1})_{lk} \, (M_{a_2})_{km}\, \chi_m$ is also a modular invariant. $N$ ensures the correct normalization of the vacuum. If two simple currents are relatively local, that is $Q^{(a_1)}(J_{a_2}) = 0$, the matrices $M_1$ and $M_2$ commute.

\subsection{Review of Gepner construction}
For $c < 3$ one finds only a discrete set of unitary $N = 2$ superconformal field theories (SCFTs), called minimal models, whose central charge
\eq{
c = {3 k \over k  + 2}
}
is parametrized by the level $k = 1,2,\dots\; $.
The primaries are labeled by three integer quantum numbers $(l, m , s)$ in the range
\eq{
l = 0, \dots k , \qquad m = -k-1, -k, \dots, k+2, \qquad s = -1,0,1,2 \,,
}
where $l+ m +s$ must be even. Additionally one needs to impose the identifications 
\eq{ \label{identification}
(l,m,s) \sim (k-l, m+k+2, s+2) , \qquad s \sim s+4, \qquad m \sim 2(k+2)\,.
}
For $s = 0,2 $ the state is in the Neveu-Schwarz (NS) sector, for $s = -1,1$ the state is in the Ramond (R) sector. To compute the conformal dimension and the charge one needs to bring $(l,m,s)$ into the standard range $|m - s| \leq l$ using \eqref{identification}. Then 
\eq{
\label{dimensions}
\Delta^l_{m,s}&={l(l+2)-m^2\over 4(k+2)} + \frac{s^2 }{ 8} \,,\\
q^l_{m,s}&={m\over (k+2)}-{s\over 2}  \,.
}
Every primary with $l = 0$ is a simple current, whose fusion with another primary reads
\eq{
\phi_{ (m_1, s_1)}^0 \times \phi_{ (m_2, s_2)}^{l_2}  =  
          \phi_{ (m_1 + m_2, s_1 + s_2)} ^{l_2} \,.
}
Gepner's construction uses tensor products of these minimal models $ \bigotimes_{i=1}^r (k_i)$ as the internal CFTs of a type II compactification with $d$ compact internal directions and $D$ extended external directions. Depending on the desired internal dimensions $d= 2, 4, 6$ the central charges of the minimal models must therefore add up to $c_{\rm int} = 3,6,9$.

Using light-cone gauge to eliminate two dimensions, we have $D - 2 = 6, 4, 2$ non-compact directions and in turn $c_{\rm ext} = 9, 6, 3$. For the external CFT one takes $D - 2$ free bosons with $c_{\rm Bos} = D - 2$. Their superpartners are $D - 2$ free fermions transforming in the vector representation of the little group $SO(D - 2)$. Their symmetry algebra is therefore the $\widehat{\mathfrak s \mathfrak o} (D - 2)_1$ Kac-Moody algebra with $c_{\rm Ferm} = {D - 2 \over 2}$. The four irreducible representations of $\widehat{\mathfrak s \mathfrak o} (D - 2)_1$ are $(c, o, s, v)$ labeled by $s_0 = -1,0,1,2$. Using $n = {D - 2 \over 2}$, their conformal weight, charge and degeneracy are
\renewcommand{\arraystretch}{1.8}
\begin{center}
\begin{tabular}{c @{\hskip 0.2in} | @{\hskip 0.3in} c @{\hskip 0.3in} cc} \label{characters}
character & $h$ & $q$ mod 2 & degeneracy  \\   \hline 
$\chi_o = {1\over 2} \Big(  \left( {\theta_3 \over \eta} \right) ^n + \left( {\theta_4 \over \eta} \right) ^n \Big) $ & $0$ & $0$ & $0$ \\
$\chi_v = {1\over 2} \Big(  \left( {\theta_3 \over \eta} \right) ^n - \left( {\theta_4 \over \eta} \right) ^n \Big) $  & ${1\over 2}$ & $1$ & $2n$ \\
$\chi_s = {1\over 2} \Big(  \left( {\theta_2 \over \eta} \right) ^n + \left( {\theta_1 \over \eta} \right) ^n \Big) $  & ${n\over 8}$ & ${n \over 2}$ & $2^{n-1}$ \\
$\chi_c = {1\over 2} \Big(  \left( {\theta_2 \over \eta} \right) ^n - \left( {\theta_1 \over \eta} \right) ^n \Big) $  & ${n\over 8}$ & ${n \over 2} - 1$ & $2^{n-1}$ \\
\end{tabular}
\end{center}
From the fusion rules
\renewcommand{\arraystretch}{1.1}
\begin{center}
\begin{tabular}{c | c c c c   c  @{\hskip 0.5in}  c | c c c c }
n odd & $o$ & $v$	 		& $s$ & $c$ & & n even & $o$ 		& $v$	 		& $s$ & $c$ \\ \cline{1-5} \cline{7-11}
$o$ & $o$ & $v$ & $s$ & $c$ &&  $o$ & $o$ & $v$ & $s$ & $c$ \\
$v$ & $v$ & $o$ & $c$ & $s$ && $v$ & $v$ & $o$ & $c$ & $s$ \\
$s$ & $s$ & $c$ & $v$ & $o$ && $s$ & $s$ & $c$ & $o$ & $v$ \\
$c$ & $c$ & $s$ & $o$ & $v$ && $c$ & $c$ & $s$ & $v$ & $o$
\end{tabular}
\end{center}
one sees that all primaries are simple currents.
To summarize, a state in a Gepner model reads
\eq{
(l_{1} \; m_1 \; s_1) \ldots (l_r \; m_r \; s_r ) (s_0) \;\;  \in \;\; \bigotimes_{i=1}^r (k_i) \otimes \widehat{\mathfrak s \mathfrak o} (D-2)_1 \, .
}
To construct a fully positive partition function one starts with a purely bosonic CFT with $c = 24$ which is mapped to a SCFT by the bosonic string map relating $\widehat{\mathfrak s \mathfrak o} (D-2)_1 \rightarrow \widehat{\mathfrak s \mathfrak o} (D+6)_1 \otimes (E_8)_1 $ via
\eq{
\phi_{\rm bsm} (\chi_o,\chi_v,\chi_s,\chi_c) \rightarrow (\chi_v,\chi_o,-\chi_c,-\chi_s) \otimes 1\,.
}
Notice that the difference of the conformal dimension of the characters in $\widehat{\mathfrak s \mathfrak o} (D-2)_1$ and $\widehat{\mathfrak s \mathfrak o} (D+6)_1$ is ${1 \over 2}$ such that level matched states stay level matched under the bosonic string map. We also need the relatively local simple currents
\eq{ \label{simplecurrents}
              J_{\rm GSO}&=(0\;1\; 1)\ldots (0\; 1\; 1)(s)\,, \\[0.1cm]
             J_i&=(0\;0\; 0)\ldots \underbrace{(0\;0\; 2)}_{i^{\rm
                th}}\ldots (0\; 0\; 0)(v) \,.
}
$J_{\rm GSO}$ implements the GSO-projection while the $J_i$ ensure that NS and R sectors are not mixed in a state. The total partition function is
\eq{ \label{partitionfn}
              Z_{\rm Gepner}(\tau,\ov\tau)\sim \vec\chi^{\,T}(\tau)\, M(J_{\rm GSO})\,
              \prod_{i=1}^r M(J_i) \,\vec\chi(\ov\tau)\Big\vert_{\phi^{-1}_{\rm bsm}} \, ,
}
where the bosonic string map has to be applied at the end and we neglected the contribution from the free bosons and possible normalization factors. This partition function allows us to read off the massless spectrum of the theory.

\subsection{Asymmetric Gepner Models} \label{asymmetricmodels}

In the following we analyze Gepner models when adding one or more possibly left-right asymmetric simple currents $J_1 ,\dots, J_n$ 
\eq{
\label{gepnerasym}
              Z_{\rm ACFT} (\tau,\ov \tau)\sim \vec\chi^{\,T}(\tau)\, M(J_{\rm 1})\dots M(J_{\rm n}) \,
              M(J_{\rm GSO})\,
              \prod_{i=1}^r M(J_i) \,\vec\chi(\ov\tau)\Big\vert_{\phi^{-1}_{\rm bsm}} \,.
}
The additional simple currents often break or enhance supersymmetry. As we want to find the ACFTs corresponding to supersymmetry breaking of a GSUGRA with more than 8 supercharges we often need to enhance supersymmetry towards our desired starting point. Afterwards the left-moving supersymmetry is broken by left-right asymmetric simple currents. In the partition function this requires to correctly order the enhancing and breaking simple currents
\eq{
              Z_{\rm ACFT} (\tau,\ov \tau)\sim \vec\chi^{\,T}(\tau)\, M(J_{\rm break}) \, M(J_{\rm enhance}) \,
              M(J_{\rm GSO})\,
              \prod_{i=1}^r M(J_i) \vec\chi(\ov\tau)\Big\vert_{\phi^{-1}_{\rm bsm}} .
}
Note that we sometimes use more than one breaking and enhancing simple current.

Let us give two four-dimensional examples for this procedure: The
Gepner model $\mathbf k = (1,3,3,4,8)$ has the standard ${\cal N} = {\cal N}_L + {\cal N}_R = 1 + 1$ supersymmetry. Taking the D-invariant in the last factor using the simple current
\eq{
J_{\rm enhance} = J_{\rm D} = (0 \; 0 \; 0)(0 \; 0 \; 0)(0 \; 0 \; 0)(0 \; 0 \; 0)(0 \; 10 \; 2)(o),
} 
supersymmetry is enhanced to ${\cal N} = {\cal N}_L + {\cal N}_R = 2 + 2$. This corresponds to the well known $K3\times \mathbb{T}^2$ compactification. Further left-right asymmetric simple currents can break the left moving supersymmetry down to ${\cal N}_{L} \in\{ 1,0\}$. As such we get models with total supersymmetry ${\cal N}  \in\{ 3,2\}$. 
A second example we will present later uses simple currents giving the $\mathbb{T}^6$ compactification with ${\cal N} = {\cal N}_L + {\cal N}_R = 4 + 4$. Further simple currents allow us to break the left moving supersymmetry down to ${\cal N}_L \in\{ 2,1,0\}$. The resulting models have therefore the total supersymmetry ${\cal N} \in\{ 6,5,4\}$.


\section{The landscape of ACFTs}
\label{sec_landscape}

In this section, we present the results of a scan over
eight-, six- and four-dimen\-sio\-nal ACFTs which feature at least
two supersymmetries arising from the right-moving sector.
In the framework of asymmetric simple current extensions of Gepner
models we constructed of the order of $10^{8}$   different modular invariant
partition functions and evaluated their  massless spectra.
It turned out that they all fall into a few different classes,
that provide a natural classification scheme for all these models.
Remarkably, even though the construction is rather abstract and
CFT based, these classes show some nice patterns that we 
make an attempt to understand from a broader perspective, i.e.
by utilizing relations following from  dimensional reduction,
gauged supergravity and  toroidal orbifolds. Before we present
the results of our ACFT landscape study, let us introduce our
classification scheme and a few  structures that will become
important.


\subsection{Classification scheme}

Our starting points are
Gepner models corresponding to $\mathbb T^2$, $K3$ and $K3\times \mathbb T^2$ 
compactifications, and 
we introduce a classification scheme $^D \mathfrak{N}_{[{\cal
    N}_L,{\cal N}_R]}$
where $D$ denotes the number of uncompactified dimensions
and ${\cal N}_{L/R}$ the number of supersymmetries arising in the ACFT from
the left and right moving sector, respectively. As mentioned, 
we only consider models with ${\cal N}_R\ge 2$. The number of
supercharges is then given by 
$Q=2^{D\over 2}\, ({\cal N}_L+{\cal N}_R)$.

Moreover, we present
the massless spectrum for some representative  ACFTs by their raw data, i.e.
we provide for a massless sector the number of massless states
transforming in the four representations $(v,s,c,o)$ of the 
left- and right-moving little groups $SO(D-2)$. Each such sector fills
out massless supermultiplets of  ${\cal N}={\cal N}_L+{\cal N}_R$ supersymmetry
in $D$-dimensions. Thus, we present the data in the form
\eq{
\arraycolsep2pt
^D \mathfrak{N}_{[{\cal N}_L,{\cal N}_R]}:\  \left\{\begin{array}{ccc@{\hspace{20pt}}l}
(n^{(0)}_v,n^{(0)}_s,n^{(0)}_c,n^{(0)}_o)_L&\otimes&
(n^{(0)}_v,n^{(0)}_s,n^{(0)}_c,n^{(0)}_o)_R\ &
{\rm supermultipl.}\\[0.2cm]
(n^{(c)}_v,n^{(c)}_s,n^{(c)}_c,n^{(c)}_o)_L&\otimes&
(n^{(c)}_v,n^{(c)}_s,n^{(c)}_c,n^{(c)}_o)_R &
{\rm supermultipl.}
\\
\ldots && \ldots & \ldots
\end{array} \right.
\nonumber
}
where, as indicated by the superscript, we state the results from the charged and the vacuum sector separately. Note that for each class, we have found a large number of concrete
realizations, while here we only present some typical
representatives.


\subsection{Super Higgs effect}
\label{sec_shiggs}

To order  the classes of models we found, we 
use some non-CFT structures.
In general the models are left-right asymmetric, i.e
they do not correspond to purely geometric 
backgrounds. In particular, non-geometric NS-NS flux compactifications are in general
expected to be 
described by ACFTs. Turning on such fluxes on e.g. $\mathbb T^4$, $K3$ or
$K3\times \mathbb T^2$ leads to gauged supergravity theories.
Of course it is difficult to identify directly the 
gaugings or fluxes from an ACFT model, but there are prerequisites 
for a gauging of a SUGRA theory with ${\cal N}'$ supersymmetries
to admit a Minkowski vacuum with ${\cal N}$ supersymmetries.
There must be the super Higgs mechanism at work \cite{Deser:1977uq,Cremmer:1978iv,Cremmer:1978hn,Andrianopoli:2002rm}.

Let us recall   this for the four-dimensional example  of 
${\cal N}'=8$ to ${\cal N}=6$ breaking. Ungauged 
${\cal N}'=8$ corresponds to the low energy effective field theory of
type II compactified on a  $\mathbb T^6$. All massless states fit into the
supergravity multiplet (see appendix~\ref{app_4} for our notation and more details)
\eq{
  \arraycolsep2pt
  \renewcommand{\arraystretch}{1.5}
  \begin{array}{l@{\hspace{30pt}}lcll}
  \mbox{massless} &\mathcal G_{(8)} 
  &= &  1\cdot [2] + 8\cdot [\tfrac{3}{2}] + 28\cdot [1] + 56\cdot [\tfrac{1}{2}] + 70\cdot [0]\\
  &&=& (2)_{\rm b} + (16)_{\rm f}   +(56)_{\rm b}  + (112)_{\rm f} + (70)_{\rm b} \,.
  \end{array}
}
The numbers in the second line denote the number of on-shell degrees
of freedom. If there exists a flux that breaks  ${\cal N}'=8$ to
${\cal N}=6$, two gravitinos must become massive and  become
part of a massive spin-$3/2$ supermultiplet of ${\cal N}=6$ supergravity.
Moreover, the remaining degrees of freedom must fit into massless
${\cal N}=6$ supermultiplets.
The massless supergravity multiplet of ${\cal N}=6$ reads
\eq{
  \arraycolsep2pt
  \renewcommand{\arraystretch}{1.5}
  \begin{array}{l@{\hspace{30pt}}lcll}
  \mbox{massless} &\mathcal G_{(6)} 
  &= &  1\cdot [2] + 6\cdot [\tfrac{3}{2}] + 16\cdot [1] + 26\cdot [\tfrac{1}{2}] + 30\cdot [0]\\
  &&=& (2)_{\rm b} + (12)_{\rm f}   +(32)_{\rm b}  + (52)_{\rm f} + (30)_{\rm b} \,,
  \end{array}
}
and the massive spin-$3/2$ supermultiplet is given by
\eq{
  \arraycolsep2pt
  \renewcommand{\arraystretch}{1.5}
  \begin{array}{l@{\hspace{30pt}}lcll}
  \mbox{massive} &\overline{\mathcal S}_{(6)} 
  &= &  2\cdot \Big(\,  [\tfrac{3}{2}] + 6\cdot [1] + 14\cdot [\tfrac{1}{2}] + 14\cdot [0]\,\Big)\\
  &&=& (8)_{\rm f}   +(36)_{\rm b}  + (56)_{\rm f} + (28)_{\rm b} \,.
  \end{array}
}
This is a ${1\over 2}$-BPS short multiplet so that for CPT invariance
it comes in a pair. 
Thus we see that the number of bosonic and fermionic degrees of
freedom perfectly match to allow that the massless ${\cal N}'=8$
gravity multiplet splits into the 
${\cal N}=6$ gravity multiplet plus a pair of massive spin-$3/2$ supermultiplets.
As a consequence kinematically the super Higgs effect is possible,
which is a necessary condition for the existence of an ${\cal N}=6$  Minkowski 
minimum in ${\cal N}'=8$ GSUGRA. 
When analyzing the ACFT results, we will often employ analogous 
super Higgs effects and its predictions on the remaining massless spectrum.


\subsection{Asymmetric $(-1)^{F_L}$  shift orbifolds}
 \label{sec_asymshift}

However, not all ACFT models in our search will admit  an interpretation in terms
of a super Higgs effect.
Even though Ramond-Ramond fluxes are expected not to be present in the ACFTs,
we find that  asymmetric orbifolds involving   $(-1)^{F_L}$ can be realized
via simple currents. Here $F_L$ denotes the left-moving space-time
fermion number, i.e. states in the left-moving NS-sector are even
and states in the left-moving R-sector are odd. Such models were
of interest for the appearance of perturbative non-abelian gauge
symmetries for the type II superstring theories \cite{Dixon:1987yp,Bluhm:1988mh}. 
Let us present a simple example in 8D that will also appear among the
ACFTs models.

We start with type IIB compactified on the rectangular $\mathbb T^2$ at
self-dual radii $r_i=\sqrt{\alpha'}$, with $i=1,2$,  for the two circles.
The partition function for the simple toroidal compactification
can then be written as
\eq{
  Z_{\mathbb T^2}=(V_8-S_8)(\tau)\, (\ov V_8-\ov S_8)(\ov\tau)\,
  \Lambda_{\vec m,\vec n}^{(2)}(\tau,\ov\tau) \,,
}
with the contribution from the Kaluza-Klein and winding modes (at $r_i=\sqrt{\alpha'}$)
\eq{
  \Lambda_{\vec m,\vec n}^{(2)}(\tau,\ov\tau)=\sum_{\vec m,\vec n\in \mathbb Z^2}
     q^{{1\over 4}\sum_i (m_i-n_i)^2}\,  \ov q^{{1\over 4}\sum_i (m_i+n_i)^2}\,.
}
Here $O_8,V_8,S_8,C_8$ denote the characters of the four conjugacy classes
of the $SO(8)_1$ Kac-Moody algebra from page \pageref{characters} and we have skipped the contribution
$1/|\eta|^{16}$ from the eight bosons.
Now we define an asymmetric orbifold
\eq{
\label{asym8d}
  {\cal A}_8={\mathbb T^2\over (-1)^{F_L}\, S\, W} \,.
}  
The orbifold projection $(-1)^{F_L}$ eliminates all states from the left-moving Ramond sector.
Moreover, $S$ denotes the momentum shift and $W$ the
winding shift along both directions of $\mathbb T^2$.
These two act on the momentum- and winding-modes as
\eq{
  S: (-1)^{\sum_i m_i}=: (-1)^{\vec m}\,,\qquad
  W: (-1)^{\sum_i n_i}=: (-1)^{\vec n}
\,.}
It is straightforward to compute the partition function
of this asymmetric orbifold as\footnote{We use the notation of
  \cite{Angelantonj:2002ct}.}
\eq{
  \label{partif}
     Z_{\rm ACFT}={1\over 2}\,\bigg[ \quad&(V_8-S_8)(\ov V_8-\ov S_8)\Lambda_{\vec m,\vec n}^{(2)}\\
+& (V_8-S_8)(\ov V_8+\ov S_8) (-1)^{\vec m +\vec n}\, \Lambda_{\vec m,\vec n}^{(2)}\\[8pt]
+& (V_8-S_8)(\ov O_8-\ov C_8) \Lambda_{\vec m+\vec{{1\over 2}},\vec n+\vec{{1\over 2}}}^{(2)}\\
+& (V_8-S_8)(\ov O_8+\ov C_8) (-1)^{\vec m +\vec n}\,\Lambda_{\vec m+\vec{{1\over 2}},\vec n+\vec{{1\over 2}}}^{(2)}\,
\bigg] \,.
 }     
The first two lines correspond to the untwisted sector and the last
two to the twisted
sector. Note that taking a momentum- or winding-shift  along
a single $S^1$, the corresponding partition function does not satisfy
the level matching condition.

The resulting massless spectrum can be read of from \eqref{partif}.
In the untwisted sector there are  $64$ bosonic and fermionic modes 
that combine into the ${\cal N}=1$ supergravity multiplet  plus
two vectormultiplets (see appendix~\ref{app_8} for details).
From the twisted sector, $V_8\, \ov O_8$ can combine
with states from
\eq{
  \Lambda_{\vec m+\vec{{1\over 2}},\vec m+\vec{{1\over 2}}}^{(2)}=
  q^0\, \sum_{\vec m} \ov q^{{1\over 4} \sum_i (2m_i+1)^2}
  }
to form  a level matched massless state. Namely, the four combinations
$m_1,m_2\in\{0,-1\}$ give rise to four ${\cal N}=1$ vectormultiplets.
These four states provide the $W$-bosons of an $SU(2)\times SU(2)$
non-abelian gauge group. Its  Coloumb-branch corresponds to
changing the two radii of the $\mathbb T^2$. 

Such a construction can be generalized to 6D and 4D, where one also
combines $(-1)^{F_L}$ with a left-moving shift of the Narain lattice
of the tori. Such models have been classified in 4D in \cite{Dixon:1987yp} with the
result that, starting with the $D_6$-lattice, one gets a model
with $[0,4]$ supersymmetry and maximal gauge symmetry $SU(2)^6$.
Analogously, in 6D one gets $[0,2]$ supersymmetry and a  maximal
$SU(2)^4$.
As we will
see, some of the models obtained in our scan can be interpreted as such asymmetric
shift orbifolds involving $(-1)^{F_L}$. We denote this class as ${\cal A}_d$.
Due to the appearance of $(-1)^{F_L}$, these models do not  correspond to 
minima of any GSUGRA theory with only NS-NS fluxes.

What is also a common feature for these exactly solvable CFTs is that
they correspond to special points with extra gauge enhancement.
Going   to the Higgs or Coulomb branch moves the model back to its
generic locus. Working with type II models,  such extra enhancements
can only arise from the NS-NS sector. The gauge and matter fields
appearing in the  R-R sector are always abelian and uncharged.


\subsection{ACFTs  in $D=8$}

Let us now present the results of our scan 
over 
asymmetric simple current extensions  in
the $\mathbf k\in\{(1,1,1), (2,2), (1,4)\}$ Gepner models with $c=3$.
Due to the simplicity of these models we could check more than $10^8$ simple current configurations but nevertheless we found only two different massless spectra.


\subsubsection*{The class $^8\mathfrak{N}_{[1,1]}$}

The first class  corresponds to the dimensional reduction
of type IIA/B on $\mathbb T^2$. This gives ${\cal N}=2$ supersymmetry
in eight dimensions with 32 supercharges and the ACFT data
are
\eq{
^8\mathfrak{N}_{[1,1]}:\  \begin{cases}
(1,1,1,2)_L\otimes
(1,1,1,2)_R &\hspace{10pt} {\cal G}_{(2)} \,.
\end{cases} 
}
Recall that the brackets count the number of states transforming in the $(v, s,c,o)$ representation
of the little group.
The massless spectrum only consists of the ${\cal N}=2$ supergravity multiplet
with bosonic field content
\eq{
\mathcal G_{(2)} = 1\cdot [2] + 2\cdot [\tfrac{3}{2}]   +6\cdot [1] + 6\cdot [\tfrac{1}{2}] + 7\cdot [0] +1\cdot [{\rm t}_3] \,.
} 
For our notation and more details on the multiplet structure in eight dimensions see appendix~\ref{app_8}.


\subsubsection*{The class $^8\mathfrak{N}_{[0,1]}$}

The second class,  $^8\mathfrak{N}_{[0,1]}$,  only appears for the $\mathbf k=(2,2)$ model for instance
with the  additional asymmetric simply current 
\eq{ \label{M8D}
              J=(0,2,2)(0,-2,2)(v) 
}
and has ${\cal N}=1$ supersymmetry.
The massless states arise in the ACFT as
\eq{
^8\mathfrak{N}_{[0,1]}:\  \begin{cases}
(1,0,0,6)_L\otimes
(1,1,1,2)_R & \hspace{10pt}{\cal G}_{(1)} + 6\cdot {\cal V}_{(1)}\,.
\end{cases} 
}
The spectrum fits into the supergravity multiplet with
field content
\eq{
  \label{nonegrav}
\mathcal G_{(1)} =
1\cdot [2] + 1\cdot [\tfrac{3}{2}]   +2\cdot [1] + 1\cdot [\tfrac{1}{2}] + 1\cdot [0] +1\cdot [{\rm t}_2] \,,
}  
and six vectormultiplets  with
\eq{
  \label{nonevec}
\mathcal V_{(1)}= 1\cdot [1] + 1\cdot [\tfrac{1}{2}] + 2\cdot [0] \,.
  }
Note that there are no massless states from a left-moving Ramond sector.
A closer look at these massless states reveals that 
these six vectors form the non-abelian gauge  group $SU(2)\times
SU(2)$. 
Recalling that
the $\mathbf k=(2,2)$ Gepner model corresponds to the rectangular $\mathbb T^2$ at
self-dual radii $r_i=\sqrt{\alpha'}$ with $i=1,2$, 
this model is nothing else than the asymmetric
$(-1)^{F_L}$ shift orbifold ${\cal A}_8$ discussed  in
section \ref{sec_asymshift}.


\subsubsection*{Summary}

In Table \ref{table8D} we summarize the ACFTs encountered in eight dimensions.
\begin{table}[ht]
\centering
\renewcommand{\arraystretch}{1.4}
\begin{tabular}{|c|c|c|}
  \hline
   class  & spectrum    &  realization\\
 \hline\hline
$^8\mathfrak{N}_{[1,1]}$ & $\mathcal G_{(2)} $ & $\mathbb T^2$\\
$^8\mathfrak{N}_{[0,1]}$ & $\mathcal G_{(1)} +6\cdot {\cal
  V}^{SU(2)^2}_{(1)}$ & ${\cal A}_8$\\
\hline
     \end{tabular} 
    \caption{\label{table8D}  Classification of type IIB ACFTs in 8D.}
\end{table}
Since the $\mathbb T^2$ trivially does not have any three-cycles to support 
NS-NS fluxes, one does not expect to find any flux compactifications.
This is consistent with our results where the only extra model
is an orbifold that involves $(-1)^{F_L}$, which is an asymmetric
operation of the Ramond sector.

However, the super-Higgs mechanism would in principle be possible. Indeed, using equation
\eqref{rel_076} from appendix~\ref{app_8} we
can decompose the eight-dimensional $\mathcal N=2$ supergravity multiplet
into $\mathcal N=1$ multiplets as
\eq{
\mathcal G_{(2)} \;\to \; \mathcal G_{(1)} + \ov{\mathcal S}_{(1)} + 
(2-\alpha)\cdot \mathcal V_{(1)} 
+ \alpha\cdot \ov{\mathcal V}_{(1)}  \,,
}
with $\alpha=0,1,2$. Hence, depending on the precise breaking mechanism, the
spontaneously-broken $\mathcal N=1$ theory can have $0,1,2$ massless
vector multiplets in addition to the supergravity multiplet. 
But, as mentioned before, since in string theory there are no suitable NS-NS fluxes available
which would give rise to the breaking, we do not expect to find these models in our search.


\subsection{ACFTs in $D=6$}

In this section we extend our investigation 
to the case $c=6$, that corresponds
to compactifications to six space-time dimensions.
Using the framework described in section~\ref{sec_gepner}, we considered
all Gepner models with $c=6$ and added up to four
additional simple currents. In general, these simple currents
do not commute with some of the generic simple currents $J_i$ 
and $J_{\rm GSO}$. 
It is remarkable that within the millions of generated type IIB models, 
we  found very few different massless spectra.
The question therefore arises, whether these models
correspond to a flux compactification on $\mathbb T^4$ or
$K3$, respectively.


\subsubsection*{The class $^6\mathfrak{N}_{[2,2]}$}

There exists the model with maximal ${\cal N}=(2,2)$
supersymmetry, which is just the compactification of type II
on a $\mathbb T^4$.  For instance the pure Gepner model $\mathbf k=(1,1,1,2,2)$
gives the massless spectrum
\eq{
^6\mathfrak{N}_{[2,2]}:\  \begin{cases}
(1,2,2,4)_L\otimes
(1,2,2,4)_R & \hspace{10pt}{\cal G}_{(2,2)}\,.
\end{cases} 
}
All massless states fit into the ${\cal N}=(2,2)$ supergravity
multiplet, on which more details can be found in appendix~\ref{app_6}.


\subsubsection*{The classes $^6\mathfrak{N}_{[1,1]}$}

Next, we consider the Gepner model  $\mathbf k=(2,2,2,2)$. 
The field content of the ACFT is
\eq{
  \label{lab_01}
^6\mathfrak{N}_{[1,1]}({\rm B}):\  \begin{cases}
\hspace{31pt}(1,0,2,0)_L\otimes
(1,0,2,0)_R & {\cal G}_{(0,2)}+{\cal T}_{(0,2)} \,,\\[0.1cm]
20\times \big[ (0,0,1,2)_L\otimes
(0,0,1,2)_R \big] \hspace{10pt}& 20\cdot {\cal T}_{(0,2)}\,,
\end{cases} 
}
so that  the model has $21$ tensor-multiplets. Recall that the
second row displays the number of massless states in the charged sector
of the simple current extension. 
Geometrically this corresponds to a compactification of the type IIB superstring
on a $K3$-manifold.

The type IIA model can be realized by a simple current extension
of the type IIB model. For $\mathbf k=(2,2,2,2)$ adding the simple current
\eq{ \label{scIIA}
J_{\rm ACFT}=(0,1,1)(0,1,1)(0,1,1)(0,1,1)(s)
} 
gives an ACFT with
\eq{
^6\mathfrak{N}_{[1,1]}({\rm A}):\  \begin{cases}
\hspace{31pt}(1,2,0,0)_L\otimes
(1,0,2,0)_R & {\cal G}_{(1,1)}\,,\\[0.1cm]
20\times \big[ (0,1,0,2)_L\otimes
(0,0,1,2)_R \big] \hspace{10pt} & 20\cdot {\cal V}_{(1,1)}\,.
\end{cases} 
}
Note the change of the $SO(4)$ representations  $c\to s$ in the
left-moving sector as compared to \eqref{lab_01}. Thus, we get a  non-chiral ${\cal N}=(1,1)$
supergravity theory with the massless spectrum of type IIA
on $K3$. Recalling that IIA = IIB/$(-1)^{F_L}$, the simple current $J_{\rm ACFT}$ is analogous
to the quotient by $(-1)^{F_L}$.


\subsubsection*{The class $^6\mathfrak{N}_{[0,2]}$}

Another model with 16 supercharges can be obtained by compactifying the
asymmetric 8D model $^8\mathfrak{N}_{[0,1]}$ on a two-torus
$\mathbb T^2$.  This can be realized as an ACFT by taking the Gepner model
$\mathbf k=(1,1,1,2,2)$ and adding the simple current
$J_{\rm ACFT}=(0,0,0)^3 (0,2,2)(0,-2,2)(v)$ similar to \eqref{M8D}. 
All states are coming from the vacuum sector
with all left-moving Ramond states projected out
 \eq{
^6\mathfrak{N}_{[0,2]}:\  \begin{cases}
(1,0,0,8)_L\otimes
(1,2,2,4)_R \hspace{10pt} & {\cal G}_{(1,1)}+8\cdot  {\cal V}_{(1,1)}\,.
\end{cases} 
}
The eight vectors transform in the gauge group $SU(2)\times
SU(2)\times U(1)^2$. Working with the ${\mathbf k}=(2,2,2,2)$ Gepner
model, we also found models with $n_V=4,8,12$ vectors. Note that
$n_V=12$ is the maximal gauge enhancement for the model ${\cal A}_6$
with gauge group $SU(2)^4$ and $n_V=4$ the minimal gauge group
$U(1)^4$. Thus, we can consider all models found in this class as
lying on the Coulomb-branch of ${\cal A}_6$.


\subsubsection*{The class $^6\mathfrak{N}_{[0,1]}$}

Now we come to the most interesting case,
i.e. models for which the asymmetric simple current
leads to minimal  chiral ${\cal N}=(0,1)$ supersymmetry.
Remarkably, we only found one such class. One representative model
is the $\mathbf k=(2,2,2,2)$ Gepner model extended by one of the simple currents $J_{\rm ACFT}=(0,0,0)^2 (0,1,1)^2(v)$ or $J_{\rm ACFT}=(0,0,0)^2 (0,2,2)(0,-2,2)(c)$. Both simple currents yield the same model and have a similar form as \eqref{4DN=2} or \eqref{M8D} that both implemented a $(-1)^{F_L}$ action. Beyond the vacuum orbit the model features three additional types of
charged orbits
\eq{ \label{6D1,0}
^6 \mathfrak{N}_{[0,1]}:\  \begin{cases}
\hspace{31pt}(1,0,0,0)_L\otimes
(1,0,2,0)_R & {\cal G}_{(0,1)}+{\cal T}_{(0,1)} \,,\\[0.1cm]
\hspace{5.5pt}8\times \big[ (0,1,0,0)_L\otimes
(0,1,0,2)_R \big] & 8\cdot {\cal T}_{(0,1)} \,,
\\[0.1cm]
\hspace{5.5pt}8\times \big[ (0,0,1,0)_L\otimes
(0,1,0,2)_R \big] & 8\cdot {\cal V}_{(0,1)} \,,
\\[0.1cm]
20\times \big[ (0,0,0,2)_L\otimes
(0,1,0,2)_R \big] \hspace{10pt} & 20\cdot {\cal H}_{(0,1)} \,.
\end{cases} 
}
Thus we have ${\cal  N}=(0,1)$ supersymmetry 
with the additional massless spectrum of tensor-, vector- and hypermultiplets
\eq{
\label{theoneandonly}
                 n_T=1+8=9\,,\qquad n_V=8\,,\qquad n_H=20\,.
}
This spectrum satisfies the anomaly cancellation condition $n_H-n_V+29
n_T=273$. 
From the ACFT analysis,  we find an extension of this model with
additional vectors and hypers, $n_V=8+n$ and  $n_H = 20 + n$ with
$n$ up to four. 
The additional hypers arise from the charged sector similar as in \eqref{6D1,0}, while the
additional vectors come from scalars in a slightly modified vacuum sector
\eq{
(1,0,0,n)_L \otimes (1,0,2,0)_R  \qquad {\cal G}_{(0,1)} + {\cal T}_{(0,1)} + n \cdot {\cal V}_{(0,1)}.
} 
For a simple example we checked that the CFT three-point functions
$\langle  v \psi \ov\psi   \rangle$ satisfied all selection rules, indicating
that the matter field $\psi$ 
carries a non-trivial $U(1)$ charge under the abelian extra gauge fields.
Therefore, these additional pairs of  (vector + hyper) can be made
massive via Higgsing.


\subsubsection*{Comments on $^6\mathfrak{N}_{[0,1]}$}

Let us comment on the reason why the class of ${\cal N}=(0,1)$ ACFTs in 
six dimensions is so restricted.

\paragraph{Anomalies} First, we observe that anomaly cancellation in addition to some
generic input from the ACFT construction
allows to restrict ${\cal  N}=(0,1)$ ACFTs considerably. Say, we have found a type IIB model of this kind with massless
spectrum $(n^B_{T, {\rm RR}} +1 \, , \, n^B_{V, {\rm RR}} + n^B_{V, {\rm NSNS}}\, , \, n^B_{H, {\rm RR}})$ where we indicated the RR and NS-NS sector in the subscript.\footnote{Notice that the indicated structure of the multiplets is completely determined by the ACFT construction. The NS-NS multiplets can only come from the vacuum sector which, due to the ${\cal N}=(0,1)$ worldsheet supersymmetry, is given by $(1,0,0,n)\otimes (1,0,2,0)$ or its charge conjugate. There is therefore always exactly one tensor multiplet from the NS-NS sector and a so far not restricted amount of vector multiplets. All other multiplets must arise from the R-R sector.} Then using the same ACFT, the corresponding type IIA model will have
the spectrum 
\eq{
(n^A_{T, {\rm RR}} +1,n^A_{V , {\rm RR}} + n^A_{V, {\rm NSNS}} ,n^A_{H, {\rm RR}})=(n^B_{V, {\rm RR}}+1 \, , \, n^B_{T, {\rm RR}} +n^B_{V, {\rm NSNS}} \,  , \, n^B_{H, {\rm RR}}) \; ,
}
i.e. tensors
and vectors from the R-R sector are exchanged while the states in the NS-NS sector match. States charged under R-R vectors do not exist, so that in  
6D one only has the gravitational  anomaly
\eq{
         {\cal A}_G=\alpha \op{\rm Tr}(R^4)+\beta\op \big({\rm Tr}\,R^2\big)^2 \,,
}
where 
\eq{     \alpha\sim 244-29\, n^{B/A}_T-n^{B/A}_H+n^{B/A}_V\,,\qquad
          \beta\sim n^{B/A}_T-8\,.
}
Requiring that both the type IIB and the correlated type IIA spectrum
cancel the irreducible anomaly immediately leads to $n^{B/A}_{T, {\rm RR}}=n^{B/A}_{V, {\rm RR}}$. Moreover, for the type II superstring there are no Chern-Simons term
so that, like for the ten-dimensional type IIB superstring, the reducible anomaly
should also cancel right away. This leads to $n^{B/A}_{T, {\rm RR}}=8$ and following from that $n^{B/A}_{V, {\rm RR}} = 8$ as well as $n_{H, {\rm RR}} = 20 + n_{V, {\rm NSNS}}$. To summarize, from this line of arguments one expects that the only
consistent ${\cal  N}=(0,1)$ spectrum arising from our ACFT
construction is the one we found.

\paragraph{Fluxes} From a supergravity point of view, a supersymmetry breaking 
to $\mathcal N=(0,1)$ could in principle be achieved by
turning on fluxes on the $K3$. However, the ACFT is expected to only
contain NS-NS fluxes, which carry three indices. Since the $K3$ only
contains two-cycles, the usual geometric and non-geometric fluxes
$H_{ijk}$, $F_{ij}{}^k$, $Q_i{}^{jk}$, $R^{ijk}$ cannot be supported.
Thus, the class $^6\mathfrak{N}_{[0,1]}$ cannot be 
considered as an NS-NS flux compactification of $K3$.

In agreement with this observation, we mention that an $\mathcal N=(0,1)$ model in six dimensions 
can not be obtained
via a spontaneous susy-breaking mechanism. For a spontaneously-broken
$(0,1)$-vacuum, gravitinos have to become massive and have to be part of a massive spin-$3/2$ multiplet. 
When decomposing for instance the $(2,2)$-theory in six dimensions into $(0,1)$-multiplets 
using the relations in \eqref{decomp_6}, we find
\eq{
  \label{rel_002}
  \mathcal G_{(2,2)} \to \mathcal G_{(0,1)} + 4 \cdot  \mathcal S^+_{(0,1)}
  + 2\cdot  \mathcal S^-_{(0,1)}
  +8\cdot  \mathcal V_{(0,1)}
  +5\cdot  \mathcal T_{(0,1)}
  + 10\cdot  \mathcal H_{(0,1)} \,.
}
However, a massive spin-$3/2$ multiplets contains one chiral and one anti-chiral gravitino, and hence
not all gravitinos in  \eqref{rel_002} can become massive. Thus, spontaneous susy-breaking from $\mathcal N=(2,2)$ to $\mathcal N=(0,1)$ in six dimensions is not possible.
For the breaking from $\mathcal N=(1,1)$ or $\mathcal N=(0,2)$ to $\mathcal N=(0,1)$ 
a similar reasoning applies. 
Using the relations in \eqref{decomp_6} we find for theories with $n_T$ tensor or $n_V$ vector multiplets
\eq{
  \mathcal G_{(0,2)} + n_T \cdot \mathcal T_{(0,2)} &\,\to \;
  \mathcal G_{(0,1)} + 2 \cdot \mathcal S^-_{(0,1)} + n_T\cdot \mathcal T_{(0,1)}
  + 2\op n_T \cdot \mathcal H_{(0,1)} \,,
  \\
  \mathcal G_{(1,1)} + n_V \cdot \mathcal V_{(1,1)} &\,\to \;
  \mathcal G_{(0,1)} + 2 \cdot \mathcal S^+_{(0,1)} + \mathcal T_{(0,1)}
  + n_V\cdot \mathcal V_{(0,1)} + 2\op n_V\cdot \mathcal H_{(0,1)} \op.
}
Again, the  gravitinos cannot  become massive and hence spontaneous super\-sym\-metry-breaking 
is not possible.

\paragraph{Orbifold realization} The above-mentioned model was discussed before in the literature \cite{Hellerman:2002ax}, where
also the following  toroidal orbifold realization was provided
\eq{
        {\rm Model\ 6D}={\mathbb T^4\over \mathbb Z_2\times \mathbb Z'_2}\,.
}
With the reflection
\eq{
            \Theta: z_i\to-z_i\,,\qquad i=1,2 \,,
}
the two discrete symmetries are given by $\Theta$ and $\Theta\, S\, (-1)^{F_L}$.    
Here $S$ denotes the momentum shift operator along a single circle $S^1$.
Note that as expected from the form of the simple currents also this orbifold involves the asymmetric operation
$(-1)^{F_L}$ that is only defined on the Ramond sector.


\subsubsection*{Summary}

To summarize, the type IIB ACFT models with $c=6$ obtained in our scan are rather restricted and 
can be characterized as shown in Table \ref{tablea6}.
\begin{table}[ht]
\centering
\renewcommand{\arraystretch}{1.4}
\begin{tabular}{|c|c|c|}
  \hline
   class & spectrum beyond SUGRA    &  realization\\
 \hline\hline
$^6\mathfrak{N}_{[2,2]}$ & $- $ & $^8\mathfrak{N}_{[1,1]}$ on $\mathbb
T^2$\\
\hline
$^6\mathfrak{N}_{[1,1]}({\rm B})$ & $21\cdot \mathcal T_{(0,2)} $ & IIB
on  $K3$\\
$^6\mathfrak{N}_{[1,1]}({\rm A})$ & $20\cdot \mathcal V_{(1,1)} $ & IIB
on $K3/(-1)^{F_L} = $ IIA on $K3$\\
$^6\mathfrak{N}_{[0,2]}$& $(4,8,12)\cdot \mathcal V_{(1,1)} $
& Coulomb-branch: ${\cal A}_6$ \\
\hline
\multirow{ 2}{*}{$^6\mathfrak{N}_{[0,1]}$} & $9\cdot \mathcal T_{(0,1)}+(8+n)\cdot \mathcal
V_{(0,1)}$ & gauge enhancement:\\
   & $+(20+n)\cdot \mathcal H_{(0,1)} $ & $\mathbb T^4/\{\Theta, \Theta S (-1)^{F_L}\}$\\
\hline
     \end{tabular} 
    \caption{\label{tablea6}  Classification of type IIB ACFTs in 6D.}
\end{table}
From this table it is clear that there is indeed no room for genuine
NS-NS flux compactifications. All asymmetric models  are realized by
orbifolds involving the projection $(-1)^{F_L}$, that only exists for the
superstring and which is of course related to the existence of a Ramond
sector.
Moreover, consistent with \cite{D'Auria:1997cz} we did not find any
model with ${\cal N}=(1,2)$ supersymmetry.


\subsection{ACFTs in $D=4$}

In this section we consider type II  Gepner models with $c=9$ and initial
${\cal N}=4$ supersymmetry,  corresponding to compactifications
on $K3\times \mathbb T^2$. Again we constructed of the order to $10^8$  individual
models\footnote{To give some numbers: Only in the $k = (2^6)$ model the stochastic search included 4.3 million
different ACFTs.}, whose massless spectra however fit into a small number of
different classes. For most of them, we can find a representative
starting with  the $\mathbf k=(1^3,2^4)$ Gepner model.
As will be presented in this section, there are models
with ${\cal N}=8,6,5,4,3,2$ supersymmetry.
Similarly to six dimensions, some of the resulting models can be 
understood as toroidal compactifications of models
in higher dimensions. However, we will also consider 
the possibility that some of the classes are flux compactifications
on $\mathbb T^6$ or $K3\times \mathbb T^2$. Note that the latter manifold 
also contains three-cycles that can support NS-NS fluxes. 
We will derive necessary constraints for such  GSUGRA models
and compare them with the ACFT results.

Let us present the classes found in the order of decreasing number of
supersymmetries. Recall that our initial model, like 
$\mathbf k=(1^3,2^4)$, has at least two right-moving supersymmetries
so that compactification on genuine Calabi-Yau three-folds are not
in our class. 


\subsubsection*{The classes $^4\mathfrak{N}_{[4,4]}$, 
$^4\mathfrak{N}_{[2,4]}$ and $^4\mathfrak{N}_{[1,4]}$}

We begin by extending the $\mathbf k=(1^3,2^4)$ Gepner model by the 
simple current 
\eq{J_{\rm ACFT}=(0,0,0)^2(0,1,1);(0,4,0)^2 (0,3,-1)^2 (o) \,,
}
which leads to an extended ${\cal N}=8$ supersymmetry.
All the massless states arise in the vacuum orbit
\eq{
^4\mathfrak{N}_{[4,4]}:\  \begin{cases}
(1,4,4,6)_L\otimes
(1,4,4,6)_R \hspace{10pt}& {\cal G}_{(8)} \,.
\end{cases} 
}
Clearly, this model corresponds to type IIB string theory compactified on a $\mathbb T^6$.

Extending instead by the simple current
\eq{
J_{\rm  ACFT}=(0,-2,0)(0,3,-1)(0,-2,2)(0,1,1)^2(0,2,2)^2 (v) \,,
}
leads to a model with ${\cal N}=6$ supersymmetry
\eq{
^4 \mathfrak{N}_{[2,4]}:\  \begin{cases}
(1,2,2,2)_L\otimes
(1,4,4,6)_R \hspace{10pt} & {\cal G}_{(6)} \,.
\end{cases} 
}
Kinematically, this model can be interpreted as a partial supersymmetry
breaking by fluxes. Indeed, as discussed in section \ref{sec_shiggs}, the massless supergravity multiplet ${\cal
  G}_{(8)}$ splits into the massless supergravity multiplet ${\cal
  G}_{(6)}$ plus a massive spin-$3/2$ multiplet. We note that
the ${\cal N}=6$ model also admits a toroidal orbifold realization
\cite{Ferrara:1989nm} as $\mathbb T^6/(\mathbb Z_2^L\, S)$,
where  $\mathbb Z_2^L$ denotes a  purely left-moving reflection of four
compact coordinates and $S$ is a $\mathbb Z_2$  shift along the
orthogonal $\mathbb T^2$. Note that this orbifold does not contain a factor
$(-1)^{F_L}$, that would transcendent a pure NS-NS flux realization.

A further breaking of the above model to ${\cal N}=5$ supersymmetry can be achieved by
the extension with two simple currents
\eq{
J_{\rm  ACFT,1}&=(0,-1,1)(0,3,-1)(0,2,0)(0,4,0)(0,2,2)(0,1,1)(0,3,-1)(v) \,, \\[4pt]
 J_{\rm ACFT,2}&=(0,-1,1)(0,2,2)(0,2,0)(0,-1,-1)(0,2,2)(0,1,1)(0,2,2)(o)\,,
}
yielding 
\eq{
^4\mathfrak{N}_{[1,4]}:\  \begin{cases}
(1,1,1,0)_L\otimes
(1,4,4,6)_R \hspace{10pt} & {\cal G}_{(5)} \,.
\end{cases} 
}
Note that for this model the maximal ${\cal  G}_{(8)}$ multiplet does not split
 into the massless supergravity multiplet ${\cal  G}_{(5)}$ plus a
 number of  massive 
spin-$3/2$ multiplets.
Thus, there is no super Higgs effect at work. However, there exist an
asymmetric orbifold realization, $\mathbb T^6/(\mathbb Z_2^L\, S,
\tilde{\mathbb Z}_2^L\, \tilde S)$, with two orthogonal shifted asymmetric reflections.


\subsubsection*{The class $^4\mathfrak{N}_{[0,4]}$}

Next we come to models featuring ${\cal N}=4$ supersymmetry in 4D.
First there exists a class that can be considered as the continuation
of the models ${\cal N}\ge 5$ just described.  The massless spectrum
arises completely from the left-moving NS-sector as
\eq{
^4\mathfrak{N}_{[0,4]}:\  \begin{cases}
(1,0,0,n)_L\otimes
(1,4,4,6)_R \hspace{10pt} & {\cal G}_{(4)}+ n \cdot {\cal V}_{(4)} \,.
\end{cases} 
}
We found a series of models
with $n_V = 0,2,4,6,8,10,14,18$. The model with $n_V=18$ can be considered
as the model ${\cal A}_4$ with maximal non-abelian gauge symmetry $SU(2)^6$.
Going to the Coulomb-branch can give the models with $n_V \ge 6$. The model with $n_V=6$ vector multiplets also arises from the super Higgs effect
\eq{
        {\cal G}_{(8)}\to  {\cal G}_{(4)}+6\cdot {\cal V}_{(4)}
}
plus two massive $1/2$-BPS  gravitino supermultiplets. Since one
massive vector-multiplet consists of two massless ones, the models
with $n_V=4,2,0$ can also be explained by flux compactifications
on $\mathbb T^6$.
Thus, the class $^4\mathfrak{N}_{[0,4]}$ can be fully explained
by two different mechanisms.


\subsubsection*{The class $^4\mathfrak{N}_{[2,2]}$}

Let us now turn to  the class $^4\mathfrak{N}_{[2,2]}$.
Their massless spectrum arises from both the vacuum and extra matter
orbits as
\eq{
^4\mathfrak{N}_{[2,2]}:\  \begin{cases}
\hspace{26.5pt}(1,2,2,2)_L\otimes
(1,2,2,2)_R \hspace{10pt} & {\cal G}_{(4)}+ 2\cdot {\cal V}_{(4)} \,,\\[0.1cm]
n\times \big[(0,1,1,2)_L\otimes
(0,1,1,2)_R\big]\hspace{10pt} & n\cdot {\cal V}_{(4)} \,.
\end{cases} 
}
Concretely, we obtained $n_V=22,14,10,6,4$. The first model is just the
$K3\times \mathbb T^2$ compactification of type IIB.

The question arises whether one can find an interpretation of the 
other four models with $n_V\in\{4,6,10,14\}$.
Since the massless spectrum is not asymmetric, one
might suspect that there exist corresponding geometric compactifications such as
toroidal orbifolds. Indeed  such toroidal orbifolds with ${\cal N}=4$ 
are given by
\eq{
                  {\rm Orb}_{n,m}=  {\mathbb T^4\times \mathbb T^2\over \mathbb Z_n\, S_m} \,,
}
where $\mathbb T^4$ is chosen such that it admits a crystallographic action of
$\mathbb Z_n$ for $n\in\{2,3,4,6\}$. Moreover, $S_m$ denotes a
momentum shift of order $m$ on $\mathbb T^2$, where $m$ is required to 
be a divisor of $n$. The computation of the resulting massless spectra
is straightforward and listed in table \ref{tableorb}.
Thus, the orbifolds provide precisely the numbers found in the ACFT
construction.
\begin{table}[ht]
\centering
\renewcommand{\arraystretch}{1.4}
\begin{tabular}{|c|r@{\hspace{4pt}}c@{\hspace{4pt}}l|c|}
  \hline
  ${\rm Orb}_{n,m}$ & \multicolumn{3}{c|}{twisted sector vectors}     &  massless spectrum\\
 \hline\hline
  $(2,2)$  & $(1,\theta)$&$=$&$(6,0)$   & ${\cal G}_{(4)}+6\cdot {\cal V}_{(4)}$ \\
   $(3,3)$  & $(1,\theta,\theta^2)$&$=$&$(4,0,0)$   & ${\cal G}_{(4)}+4\cdot 
   {\cal V}_{(4)}$ \\
$(4,2)$  & $(1,\theta,\theta^2,\theta^3)$&$=$&$(4,0,10,0)$   & ${\cal
  G}_{(4)}+14\cdot {\cal V}_{(4)}$ \\
$(6,3)$  & $(1,\theta,\theta^2,\theta^3,\theta^4,\theta^5)$&$=$&$(4,0,0,6,0,0)$   & ${\cal
  G}_{(4)}+10\cdot  {\cal V}_{(4)}$ \\
\hline
     \end{tabular} 
    \caption{\label{tableorb}  Spectra of shift orbifolds. The spectra
      of the other models ${\rm Orb}_{4,4}$, ${\rm Orb}_{6,6}$ and
      ${\rm Orb}_{6,2}$ give $n_V=\{4,4,14\}$, which are  already contained in
      the list. }
\end{table}


\subsubsection*{The class $^4\mathfrak{N}_{[1,2]}$}

We also found a class of  models featuring ${\cal N}=3$ supersymmetry
in 4D.
One representative originates from the $\mathbf k=(1^3,2^4)$ Gepner
model via extension by the simple current
\eq{
J_{\rm ACFT}=(0,3,-1)(0,1,1)(0,-2,0)(0,-1,-1)(0,-2,2)(0,4,0)(0,-3,1) (s)\,.
}
The massless spectrum reads
\eq{
^4\mathfrak{N}_{[1,2]}:\  \begin{cases}
\hspace{25.5pt} (1,1,1,0)_L\otimes
(1,2,2,2)_R \hspace{10pt} & {\cal G}_{(3)}+ {\cal V}_{(3)}\,,\\[0.1cm]
6\times \big[(0,2,0,2)_L\otimes
(0,1,1,2)_R\big] \hspace{10pt} & 6\cdot {\cal V}_{(3)}\,,\\[0.1cm]
6\times \big[(0,0,2,2)_L\otimes
(0,1,1,2)_R\big] \hspace{10pt} & 6\cdot {\cal V}_{(3)}\,.
\end{cases} 
}
Thus, besides the ${\cal N}=3$ supergravity multiplet there are 13 vectormultiplets. Our stochastic search also provided models with 
\eq{
n_V\in\{3, 7, 11, 13, 19\} \,.
}

Let us analyze whether this class could arise via the super Higgs
effect from a 4D theory with higher supersymmetry, which  would
for instance correspond to a flux compactification on $\mathbb T^4$ or $K3\times \mathbb T^2$.
Using the supermultiplet structure reviewed in appendix~\ref{app_4}, one
can straightforwardly derive the table \ref{tablesuperHiggs} for admissible super
Higgsing.
\begin{table}[ht]
\centering
\renewcommand{\arraystretch}{1.4}
\begin{tabular}{|c|c|c|}
  \hline
  $ {\cal N}'$ & ${\cal N}$    &  massless spectrum\\
 \hline\hline
  $8$  & $3$   & ${\cal G}_3+(3-2k)\cdot {\cal V}_3$ \\
  $6$  & $3$   & $-$ \\
  $5$  & $3$   & $-$ \\
 $4$  & $3$   & ${\cal G}_3+(19-2k)\cdot {\cal V}_3$ \\
\hline
     \end{tabular} 
    \caption{\label{tablesuperHiggs} Admissible super Higgs effect ${\cal
        N}'\to {\cal N}=3$ with $k\in\mathbb N_0$.}
\end{table}
Therefore, all the models we found with $n_V\le 19$ can be understood
as flux compactifications on $K3\times \mathbb T^2$.  The model with $n_V=3$
could also be interpreted as arising via super Higgs mechanism from $
{\cal N}'=8$, i.e. a flux model on $\mathbb T^6$.
Note that the ACFT data are completely consistent with an
interpretation in terms of a super Higgs effect, i.e. a Minkowski
type flux vacuum. We see the upper bound $n_V=19$ and only odd
numbers of vector multiplets.


\subsubsection*{The class $^4\mathfrak{N}_{[0,2]}$}

We now turn to the case of ${\cal N}=2$ supersymmetry in
4D.  Of course, most Gepner models directly give models of the type
$^4\mathfrak{N}_{[1,1]}$,
corresponding to compactifications of type IIB string theory on genuine Calabi-Yau
three-folds. However, here we are not interested in these cases.
Instead we search for ACFTs in the class $^4
\mathfrak{N}_{[0,2]}$, as these could arise from Minkowski minima of
flux compactifications on $K3\times  \mathbb T^2$.

We find only three different classes of models that are distinguished by
the difference of the number of vector and hypermultiplets.
The massless spectrum of the first class reads
\eq{
\label{4DN=2}
^4\mathfrak{N}_{[0,2]}({\rm A}):\  \begin{cases}
\hspace{4pt} (1,0,0,m)_L\otimes
(1,2,2,2)_R \hspace{10pt}& {\cal G}_{(2)}+ (m+1)\cdot {\cal V}_{(2)}\,, \\[0.1cm]
(0,n,n,2k)_L\otimes
(0,1,1,2)_R\hspace{10pt} & 2\op n\cdot {\cal V}_{(2)}+ k\cdot {\cal H}_{(2)} \,,
\end{cases} 
}
with $n\ge 1$. Thus, such models have $n_V=m+2n+1$ vectormultiplets and
$n_H=k$ hypermultiplets. In our search we always find $n_H=n_V+1$ with
a long list for the number of vectormultiplets
\eq{
n_V\in\{1,3,5,6,7,8,9,10,11,12,13,14,15,17,19,20,21,22,23\}\,.
}
\begin{itemize}

\item The model with $n_V=19$ can be interpreted as the 6D model 
$^6\mathfrak{N}_{[0,1]}$ compactified on $\mathbb T^2$. As in 6D, we find up to 
four additional vector/hyper pairs.

\item Let us analyze  the possibility of the super Higgs effect to occur for
${\cal N}=4\to {\cal N}=2$ in four dimensions.
Let us denote by $n^{L,S}_{3/2}$ the number of long and short
massive ${\cal N}=2$ gravitino multiplets and by
$n^{L,S}_{1}$ the number of long and short
massive ${\cal N}=2$ vector  multiplets.
The super Higgs effects leave 
\eq{
  &n_V=27-4(n^{L}_{3/2}+n^{S}_{3/2})-(n^{L}_{1}+2 n^{S}_{1}) \hspace{20pt}\mbox{and} \\[4pt]
  &n_H=20+4 n^{S}_{3/2}-n^{L}_{1} 
}
massless  vector- and hypermultiplets.
For the case of no massive short multiplets, one finds
$n_V=19-n^{L}_{1}$ and $n_H=20-n^{L}_{1}$ so that indeed $n_H-n_V=1$.
Therefore, {\em all} the models with $n_V\le 19$ are consistent with
the expectation form a super Higgs mechanism. We find it quite
remarkable that our list of ACFTs  covers (almost) all possible values
of $n_V$. We expect that the few gaps will also be filled 
by running an even more extensive search.
\end{itemize}

\noindent
Let us emphasize that precisely for the case of flux compactifications on
$K3\times \mathbb T^2$, we find an increase in the number of ACFTs, all
of them consistent with the GSUGRA predictions $n_H-n_V=1$ and
$n_V\le 19$.

\vspace{0.2cm}
The second class $^4\mathfrak{N}_{[0,2]}({\rm B})$ has a massless
spectrum of the same form \eqref{4DN=2} though obeys $n_V - n_H = 11$ with
\eq{
n_V \in \{ 13, 15, 17, 19, 21, 23\}\,.
}
Before interpreting the second class let us introduce the third class consisting of models without massless states from the left-moving
Ramond  sector. The massless spectrum has the following structure
\eq{
^4 \mathfrak{N}_{[0,2]}({\rm C}):\  \begin{cases}
\hspace{1.5pt}(1,0,0,m)_L\otimes
(1,2,2,2)_R \hspace{10pt} & {\cal G}_{(2)}+ (m+1)\cdot {\cal V}_{(2)} \,, \\[0.1cm]
(0,0,0,2k)_L\otimes
(0,1,1,2)_R \hspace{10pt} &  k\cdot {\cal H}_{(2)} \,.
\end{cases} 
}
All models obtained in our scan corresponding to this class satisfy $n_H-n_V=13$ with \eq{
n_V\in\{3,4,5,7,8,9,10,11\}\,.
}
The even models were very rare so that we are confident that a more-extensive scan would fill the gaps.
\begin{itemize}
\item For $n_V\le 7$ the models can arise via the super Higgs effect from
${\cal N}=4$ through $n_{3/2}^S=0$ and $n_{1}^S=6$. In this case, we
find $n_V=7-n^{L}_{1}$ and $n_H=20-n^{L}_{1}$. 

\item 
Alternatively, the model with $n_V=7$ also results from compactifying
the asymmetric 8D model $^8\mathfrak{N}_{[0,1]}={\cal A}_8$ on a
$K3$ manifold. 
One way to see this is to consider the orbifold realization
\eq{
  \label{lab_02}
     {\mathbb T^4\times \mathbb T^2\over \{\mathbb Z_2, (-1)^{F_L} SW\}} \,,
}
where the $\mathbb Z_2$ acts as a reflection $\theta:x_i\to -x_i$ on the
four coordinates of $\mathbb T^4$. 
Having the structure of a $\mathbb Z_2\times \mathbb Z_2$ orbifold, one can introduce 
a discrete torsion $\epsilon=\pm 1$.
In Table \ref{tabledk3} we display the resulting massless spectra for these
two choices, indicating the various twisted sector
contributions. As one can see the choice of $\epsilon = 1$ gives a spectrum from the third class while $\epsilon = -1$ fits into the second class of ACFT models.

Note that for the $\epsilon=+1$ model there are no massless modes
from the R-R sector, both in the untwisted and twisted sectors.
On the other hand, for  the $\epsilon=-1$ model in the $\theta$-twisted sector the
NS-NS sector is projected out and the R-R sector is kept.

Therefore the last two ACFT classes above can be interpreted as the Higgs 
branches of these two models. We note that, as observed before in 6D,
up to  four additional vector/hyper pairs can become massless. In the
moment, we cannot say whether this is a strong upper bound that maybe
has a natural interpretation.

\end{itemize}

\begin{table}[ht]
\centering
\renewcommand{\arraystretch}{1.4}
\begin{tabular}{|c|c|c|}
  \hline
  sector  &  $\epsilon=+1$     &    $\epsilon=-1$\\
 \hline\hline
untwisted & ${\cal G}_{(2)} + 3\cdot {\cal V}_{(2)} + 4\cdot {\cal H}_{(2)}$ &
${\cal G}_{(2)} + 3\cdot {\cal V}_{(2)} + 4\cdot {\cal H}_{(2)}$\\
$\theta$ twisted & $16\cdot {\cal H}_{(2)}$ & $16\cdot {\cal V}_{(2)}$  \\
$(-1)^{F_L} SW$ twisted & $4\cdot {\cal V}_{(2)}$ & $4\cdot {\cal H}_{(2)}$
\\
\hline
total & ${\cal G}_{(2)} + 7\cdot {\cal V}_{(2)} + 20\cdot {\cal H}_{(2)}$ &
${\cal G}_{(2)} + 19\cdot {\cal V}_{(2)} + 8\cdot {\cal H}_{(2)}$\\
\hline
     \end{tabular} 
    \caption{\label{tabledk3} Massless spectra for the orbifold \eqref{lab_02} with and without 
    discrete torsion.}
 
\end{table}


\subsubsection*{Summary }

To summarize, the four-dimensional type IIB ACFT models with $c=9$ are rather restricted.
In particular, they can be characterized according 
to the classification shown in Table \ref{tablec}.

\begin{table}[ht]
\centering
\tabcolsep5pt
\renewcommand{\arraystretch}{1.4}
\hspace*{-14.01pt}\begin{tabular}{|c|c|c|}
  \hline
   class & spectrum beyond SUGRA    &  realization\\
 \hline\hline
$^4\mathfrak{N}_{[4,4]}$ & $- $ & type IIB on $\mathbb T^6$\\
$^4\mathfrak{N}_{[2,4]}$ & $- $ & sHiggs of
$^4 \mathfrak{N}_{[4,4]}$ \\
$^4 \mathfrak{N}_{[1,4]}$ & $-$ & $-$ \\
\hline
\multirow{2}{*}{$^4  \mathfrak{N}_{[0,4]}$} & $(0,2,4,6)\cdot
\mathcal V_{(4)} $ & sHiggs of $^4\mathfrak{N}_{[4,4]}$ \\
   & $(6,8,10,14,18)\cdot
\mathcal V_{(4)} $ & Coulomb branch: ${\cal A}_4$\\[6pt]
$^4 \mathfrak{N}_{[2,2]}({\rm A})$ & $22\cdot \mathcal
V_{(4)} $ & type IIB
on  $K3\times \mathbb T^2$\\[6pt]
$^4  \mathfrak{N}_{[2,2]}({\rm B})$ & $(4,6,10,14)\cdot \mathcal
V_{(4)} $ & shift orbifolds ${\rm Orb}_{n,m}$\\
\hline
$^4 \mathfrak{N}_{[1,2]}$ & $(3,7,11,13,19)\cdot \mathcal V_{(3)}
$ & sHiggs of $^4\mathfrak{N}_{[2,2]}({\rm A})$
\\[6pt]
\multirow{2}{*}{$^4 \mathfrak{N}_{[0,2]}({\rm A})$}  & $(1,\ldots,19)\cdot \mathcal V_{(2)}+ 
(2,\ldots,20)\cdot \mathcal H_{(2)}$ & sHiggs of $^4\mathfrak{N}_{[2,2]}({\rm A})$\\
  & $(19,\dots , 23 )\cdot \mathcal V_{(2)} +
(20 ,\dots, 24) \cdot \mathcal H_{(2)}$ &  $^6\mathfrak{N}_{[0,1]}$
on $\mathbb T^2$
\\[6pt]
\multirow{2}{*}{$^4 \mathfrak{N}_{[0,2]}({\rm B})$}  & $(13, 15, 17 , 19) \cdot \mathcal V_{(2)} + 
(2,4, 6, 8)\cdot \mathcal H_{(2)}$ & Higgs chain:  $^8\mathfrak{N}_{[0,1]}$ on $K3_{\epsilon=-1}$ \\
& $(21, 23) \cdot \mathcal V_{(2)} +
(10,12)  \cdot \mathcal H_{(2)}$ & gauge enhancement\\[6pt]
\multirow{2}{*} {$^4 \mathfrak{N}_{[0,2]}({\rm C})$}
& $ (3,4,5,7) \cdot \mathcal V_{(2)} +
(16,17,18,20)  \cdot \mathcal H_{(2)}$ &Higgs chain:  $^8\mathfrak{N}_{[0,1]}$ on $K3_{\epsilon = +1}$\\
& $(8,9,10,11) \cdot \mathcal V_{(2)} +
(21,22,23,24) \cdot \mathcal H_{(2)}$ & gauge enhancement\\
\hline
     \end{tabular} 
    \caption{\label{tablec}  Classification of type IIB ACFTs in 4D.}
\end{table}


\section{Conclusions}
\label{sec_concl}

In this paper we have presented the results of an extensive stochastic computer search for 
simple current extended asymmetric Gepner models with at least
eight  supercharges in the right-moving sector. Our main motivation was to continue
the analysis of four-dimensional ${\cal N}=1$ ACFTs from \cite{Blumenhagen:2016axv} in a simpler
setting, where a clearer picture could arise. The main result of
\cite{Blumenhagen:2016axv} was a proposal for the identification of certain ACFTs 
as Minkowski minima of gauged supergravities in specified Calabi-Yau three-folds.
This proposal had some ambiguities related to the existence of a scalar potential
in four-dimensional $\mathcal N=1$ theories. 
On the other hand, for models with eight supercharges considered in this paper, the generation of mass terms
and scalar potentials is much more restricted, as  often gauge
fields and scalars reside in the same supermultiplet.

We considered ACFTs in $D=8,6,4$ dimensions and were able to characterize
our results by just a few classes. In 8D we only 
found two models, where one was  just the $\mathbb T^2$ compactification of
type IIB string theory in 10D. The second model  could be identified with an asymmetric
orbifold that involved momentum/winding shifts and $(-1)^{F_L}$. 
Moreover, rather reminiscent of the heterotic string, there appeared
a non-abelian gauge symmetry. This model makes it very clear that
there exist asymmetric ACFTs that cannot correspond to NS-NS flux
compactifications, as they involve the Ramond sector.
 
In 6D, the number of different ACFTs only increased slightly,  all of
them again being describable via  asymmetric
orbifolds involving $(-1)^{F_L}$. As we argued, there was no sign of the
super Higgs mechanism, a prerequisite of spontaneous supersymmetry
breaking via fluxes or gaugings, respectively. And indeed, since there are no
three-cycles on $K3$, NS-NS fluxes cannot be supported and hence spontaneous
partial susy breaking is not expected.

In 4D the landscape became richer but still the models could
be classified according to their supersymmetry and the four
categories: dimensional reduction, asymmetric $(-1)^{F_L}$ shift orbifolds, 
special CFT gauge enhancement and, last but not least, the 
super Higgs mechanism. The latter was expected via flux
compactifications on $\mathbb T^6$ and $K3\times \mathbb T^2$ and indeed
for ${\cal N}=3$ and ${\cal N}=2$ supersymmetry two chains
of models were obtained that precisely fit into this scheme.
Of course, this does not  yet prove that ${\cal N}=4$ gauged supergravity
really admits these Minkowski vacua by concrete choices of gaugings,
but it provides compelling evidence. In fact, at least for certain
models there  could exist also an asymmetric orbifold realization (not
involving $(-1)^{F_L}$), though this does not exclude an
interpretation in terms GSUGRA. As in the fee fermion construction
\cite{Ferrara:1989nm}, we found models with  ${\cal N}=8,6,5,4,3,2$
supersymmetry.  Of course, in view of the ``landscape versus swampland'' question,
it would be desirable to identify the
precise relation between the ACFT data and the concrete gaugings or
fluxes. This is not an easy question and is beyond the scope of 
this paper.

To summarize, the landscape of asymmetric Gepner models is rich but still
features a clear structure. Of course, these models do not
cover all parts of the string landscape. In particular, we do not
expect to find models with R-R fluxes turned on. Moreover, 
probably not all modular invariant partition functions can be reached
via the simple current construction. There could well be string islands \cite{Dabholkar:1998kv}
that only a full classification of modular invariant partition
functions can reveal. 


\vskip2em
\noindent
\emph{Acknowledgments:} We would like to thank
O.~Andreev, G.~Dall'Agata and I.~Garc{\'i}a-Etxebarria
for helpful discussions.


\clearpage
\appendix


\section{Supermultiplets}

In this appendix we review the field content of the supermultiplets
in various dimensions. Part of this information can be found in
\cite{Strathdee:1986jr}, and the full summary for four dimensions in 
\cite{Andrianopoli:2002rm}.


\subsection{Supergravity in $D=8$}
\label{app_8}

In this section we collect some information about the multiplet structure
in eight dimensions. 
The on-shell degrees of freedom of the various fields are summarized as
follows
\eq{
  \renewcommand{\arraystretch}{1.3}
  \begin{array}{l@{\hspace{6pt}}l@{\hspace{20pt}}c@{\hspace{20pt}}c}
  \multicolumn{2}{c}{\mbox{name}} & \mbox{symbol} &\mbox{on-shell d.o.f.}
  \\ \hline\hline
  \mbox{massless}&\mbox{spin~}2&[2]&20_{\rm b}
  \\
  \mbox{massless}&\mbox{spin~}$3/2$&[\tfrac{3}{2}]&40_{\rm f} 
  \\
  \mbox{massless}&\mbox{spin~}$1$&[1]&6_{\rm b}
  \\
  \mbox{massless}&\mbox{spin~}$1/2$&[\tfrac{1}{2}]&8_{\rm f} 
  \\
  \mbox{massless}&\mbox{spin~}$0$&[0]&1_{\rm b}
  \\
  \mbox{massless}&\mbox{$p$-form}&[{\rm t}_p]&\binom{6}{p}_{\rm b}\!\!
  \\[4pt] \hline
  &&&
  \\[-15pt]
  \mbox{massive}&\mbox{spin~}$3/2$&\ov{[\tfrac{3}{2}]}&48_{\rm f}
  \\
  \mbox{massive}&\mbox{spin~}$1$&\ov{[1]}&7_{\rm b}
  \\
  \mbox{massive}&\mbox{spin~}$1/2$&\ov{[\tfrac{1}{2}]}&8_{\rm f}
  \\
  \mbox{massive}&\mbox{spin~}$0$&\ov{[0]}&1_{\rm b}
  \\
  \mbox{massive}&\mbox{$p$-form}&\ov{[{\rm t}_p]}&\binom{7}{p}_{\rm b}\!\!
  \end{array}
}  
The multiplets relevant for our discussion 
are summarized in table~\ref{sum_d8}.
In terms of the field content, these multiplets satisfy the following relations
\eq{
  \label{rel_076}
  &\mathcal G_{(2)} = \mathcal G_{(1)} + \mathcal S_{(1)} + 2\cdot \mathcal V_{(1)} \,,
  \\[6pt]
  &\ov{\mathcal S}_{(1)} = \mathcal S_{(1)}\,,
  \\
  &\ov{\mathcal V}_{(1)} = \mathcal V_{(1)}\,.
}

\begin{table}[p]
\setlength{\LTcapwidth}{\linewidth}
\tabcolsep9.8pt
\renewcommand{\arraystretch}{1.4}
\begin{longtable}[c]{c|c|c|l@{\hspace{2pt}}c@{\hspace{2pt}}l}
$\mathcal N$ & spin & mass & \multicolumn{3}{l}{content}
\\ \hline\hline
$2$ & $2$ &  & $\mathcal G_{(2)}$ &$=$&
$1\cdot [2] + 2\cdot [\tfrac{3}{2}]   +6\cdot [1] + 6\cdot [\tfrac{1}{2}] + 7\cdot [0] +1\cdot [{\rm t}_3]$
\\
&&&&&
$\hspace{1pt}+\,3\cdot [{\rm t}_2]$
\\[10pt]  
$1$ & $2$ &  & $\mathcal G_{(1)}$ &$=$&
$1\cdot [2] + 1\cdot [\tfrac{3}{2}]   +2\cdot [1] + 1\cdot [\tfrac{1}{2}] + 1\cdot [0] +1\cdot [{\rm t}_2]$
\\
$1$ & $3/2$ &  & $\mathcal S_{(1)}$ &$=$&
$1\cdot [\tfrac{3}{2}]   +2\cdot [1] + 3\cdot [\tfrac{1}{2}] + 2\cdot [0] +2\cdot [{\rm t}_2] + 1\cdot [{\rm t}_3]$
\\
$1$ & $3/2$ & long & $\ov{\mathcal S}_{(1)}$ &$=$&
$1\cdot \ov{[\tfrac{3}{2}]}   +1\cdot \ov{[1]} + 2\cdot \ov{[\tfrac{1}{2}]} + 1\cdot \ov{[0]} +1\cdot \ov{[{\rm t}_2]} + 1\cdot \ov{[{\rm t}_3]}$
\\
$1$ & $1$ &  & $\mathcal V_{(1)}$ &$=$&
$1\cdot [1] + 1\cdot [\tfrac{1}{2}] + 2\cdot [0] $
\\
$1$ & $1$ & long & $\ov{\mathcal V}_{(1)}$ &$=$&
$1\cdot \ov{[1]} + 1\cdot \ov{[\tfrac{1}{2}]} + 1\cdot \ov{[0]} $
\vspace*{20pt}
\\
\caption{Supergravity multiplets in $D=8$. The first column shows the amount of
supersymmetry, the second column
indicates the maximal spin of the multiplet, the third column
specifies whether the fields are massless (no indication) or
massive (long).
\label{sum_d8}}
\end{longtable}
\end{table}


\clearpage
\subsection{Supergravity in $D=6$}
\label{app_6}

In six dimensions, the on-shell degrees of freedom of the various fields are summarized as
follows
\eq{
  \renewcommand{\arraystretch}{1.3}
  \begin{array}{l@{\hspace{6pt}}l@{\hspace{20pt}}c@{\hspace{20pt}}c}
  \multicolumn{2}{c}{\mbox{name}} & \mbox{symbol} &\mbox{on-shell d.o.f.}
  \\ \hline\hline
  \mbox{massless}&\mbox{spin~}2&[2]&9_{\rm b}
  \\
  \mbox{massless}&\mbox{spin~}$3/2$&[\tfrac{3}{2}]^{\pm}&6_{\rm f} 
  \\
  \mbox{massless}&\mbox{spin~}$1$&[1]&4_{\rm b}
  \\
  \mbox{massless}&\mbox{spin~}$1/2$&[\tfrac{1}{2}]^{\pm}&2_{\rm f} 
  \\
  \mbox{massless}&\mbox{spin~}$0$&[0]&1_{\rm b}
  \\
  \mbox{massless}&\mbox{two-form}&[{\rm t}_2]^{\pm}&3_{\rm b}
  \\[4pt] \hline
  &&&
  \\[-15pt]
  \mbox{massive}&\mbox{spin~}$3/2$&\ov{[\tfrac{3}{2}]}&16_{\rm f}
  \\
  \mbox{massive}&\mbox{spin~}$1$&\ov{[1]}&5_{\rm b}
  \\
  \mbox{massive}&\mbox{spin~}$1/2$&\ov{[\tfrac{1}{2}]}&4_{\rm f}
  \\
  \mbox{massive}&\mbox{spin~}$0$&\ov{[0]}&1_{\rm b}
  \\
  \mbox{massive}&\mbox{two-form}&\ov{[{\rm t_2}]}&10_{\rm b}
  \end{array}
}  
Note that the $\pm$ indicates the chirality of the fermionic fields, and
whether the two-tensor is self- or anti-self-dual.
The multiplets relevant for our discussion are summarized in 
table~\ref{sum_d6}.
At the level of the field content, the following relations can be
obtained
\eq{
  \label{decomp_6}
  \arraycolsep2pt
  \renewcommand{\arraystretch}{1.4}
  \begin{array}{lcl@{\hspace{-30pt}}lcl}
  \mathcal G_{(2,2)} &=& \mathcal G_{(0,2)} + 4\cdot \mathcal S_{(0,2)} + 5\cdot \mathcal T_{(0,2)} \,, \\
  \mathcal G_{(2,2)} &=& \mathcal G_{(1,1)} + 2\cdot \mathcal S^+_{(1,1)}+ 2\cdot \mathcal S^-_{(1,1)}
   + 4\cdot \mathcal V_{(1,1)} \,, \\[6pt]
  \mathcal G_{(0,2)} &=& \mathcal G_{(0,1)} + 2\cdot \mathcal S_{(0,1)}^- \,, &
  \mathcal G_{(1,1)} &=& \mathcal G_{(0,1)} + 2\cdot \mathcal S_{(0,1)}^+ + \mathcal T_{(0,1)} \,, \\
  \mathcal S_{(0,2)} &=& \mathcal S^+_{(0,1)} + 2\cdot \mathcal V_{(0,1)} \,, &
  \mathcal S^+_{(1,1)} &=& \mathcal S_{(0,1)}^+ + 2\cdot\mathcal T_{(0,1)} \,, \\
  \mathcal T_{(0,2)} &=& \mathcal T_{(0,1)} + 2\cdot \mathcal H_{(0,1)} \,, 
  &\mathcal S^-_{(1,1)} &=& \mathcal S_{(0,1)}^- + 2\cdot\mathcal V_{(0,1)}+  \mathcal H_{(0,1)} \,, \\
  &&&
  \mathcal V_{(1,1)} &=& \mathcal V_{(0,1)} + 2\cdot \mathcal H_{(0,1)} \,, \\
  \ov{\mathcal S}_{(2)} &=& \mathcal S_{(1,1)}^+ + \mathcal S_{(1,1)}^- \,, \\
  \ov{\mathcal V}_{(2)} &=& \mathcal V_{(1,1)} \,, \\[6pt]
  \ov{\mathcal S}_{(1)} &=& \mathcal S^+_{(0,1)}+\mathcal S^-_{(0,1)} 
  + 2\cdot \mathcal V_{(0,1)} + 2\cdot \mathcal T_{(0,1)} + \mathcal H_{(0,1)}\,, \\
  \ov{\mathcal V}_{(1)} &=& \mathcal V_{(0,1)} + 2\cdot \mathcal H_{(0,1)}\,. \\[-16pt]
  \end{array}
}

\begin{table}[p]
\setlength{\LTcapwidth}{\linewidth}
\tabcolsep10pt
\renewcommand{\arraystretch}{1.35}
\begin{longtable}{c|c|c|l@{\hspace{2pt}}c@{\hspace{2pt}}l}
$\mathcal N$ & spin & mass & \multicolumn{3}{l}{content}
\\ \hline\hline
$(2,2)$ & $2$ &  & $\mathcal G_{(2,2)}$ &$=$&
$1\cdot [2] + 4\cdot [\tfrac{3}{2}]^+ + 4\cdot [\tfrac{3}{2}]^- +16\cdot[1]$ \\
&&&&&$\hspace{1pt}+\,20\cdot[\tfrac{1}{2}]^++20\cdot[\tfrac{1}{2}]^-+25\cdot[0]$
\\
&&&&&
$\hspace{1pt}+\,5\cdot [{\rm t}_{2}]^++5\cdot [{\rm t}_{2}]^- $
\\[10pt]
$(0,2)$ & $2$ &  & $\mathcal G_{(0,2)}$ &$=$&
$1\cdot [2] + 4\cdot [\tfrac{3}{2}]^-   +5\cdot [{\rm t}_2]^- $
\\
$(0,2)$ & $3/2$ &  & $\mathcal S_{(0,2)}$ &$=$&
$1\cdot [\tfrac{3}{2}]^+   +4\cdot [1]+ 5\cdot[\tfrac{1}{2}]^- $
\\
$(0,2)$ & $1$ &  & $\mathcal T_{(0,2)}$ &$=$&
$1\cdot [{\rm t}_2]^+ + 4\cdot [\tfrac{1}{2}]^+   +5\cdot [0] $
\\[10pt]  
$(1,1)$ & $2$ &  & $\mathcal G_{(1,1)}$ &$=$&
$1\cdot [2] + 2\cdot [\tfrac{3}{2}]^++ 2\cdot [\tfrac{3}{2}]^- 
  +4\cdot [1] $ \\
&&&&&$\hspace{1pt}+\,2\cdot [\tfrac{1}{2}]^++2\cdot [\tfrac{1}{2}]^- + 1\cdot [0]$
  \\
&&&&&
$\hspace{1pt}+\,1\cdot[{\rm t}_2]^++1\cdot[{\rm t}_2]^-$
\\
$(1,1)$ & $3/2$ &  & $\mathcal S^{\pm}_{(1,1)}$ &$=$&
$1\cdot [\tfrac{3}{2}]^{\pm}+ 2\cdot [1] + 4\cdot [\tfrac{1}{2}]^{\pm}+ 1\cdot [\tfrac{1}{2}]^{\mp}  $ \\
&&&&&$\hspace{1pt}+\,2\cdot [0] + 2 \cdot [{\rm t}_2]^{\pm} $
\\
$(1,1)$ & $1$ &  & $\mathcal V_{(1,1)}$ &$=$&
$1\cdot [1] + 2\cdot [\tfrac{1}{2}]^++2\cdot [\tfrac{1}{2}]^-   +4\cdot [0]  $
\\[10pt]
$2$ & $3/2$ & short & $\ov{\mathcal S}_{(2)}$ &$=$&
$1\cdot \ov{[\tfrac{3}{2}]}+2\cdot \ov{ [1]} + 4\cdot \ov{[\tfrac{1}{2}]}   +2\cdot \ov{[0]}  +2\cdot \ov{[{\rm t}_2]}$
\\
$2$ & $1$ & short & $\ov{\mathcal V}_{(2)}$ &$=$&
$1\cdot \ov{ [1]} + 2\cdot \ov{[\tfrac{1}{2}]}   +3\cdot \ov{[0]}  $
\\[10pt]
$(0,1)$ & $2$ &  & $\mathcal G_{(0,1)}$ &$=$&
$1\cdot [2] + 2\cdot [\tfrac{3}{2}]^- + 1\cdot[{\rm t}_2]^-$
\\
$(0,1)$ & $3/2$ &  & $\mathcal S^+_{(0,1)}$ &$=$&
$1\cdot [\tfrac{3}{2}]^+ + 2\cdot [1]   + 1\cdot [\tfrac{1}{2}]^-$
\\
$(0,1)$ & $3/2$ &  & $\mathcal S^-_{(0,1)}$ &$=$&
$1\cdot [\tfrac{3}{2}]^- + 2\cdot [{\rm t}_2]^-$
\\
$(0,1)$ & $1$ &  & $\mathcal V_{(0,1)}$ &$=$&
$1\cdot [1] + 2\cdot [\tfrac{1}{2}]^- $
\\
$(0,1)$ & $0$ &  & $\mathcal H_{(0,1)}$ &$=$&
$1\cdot [\tfrac{1}{2}]^+ + 2\cdot[0]$
\\
$(0,1)$ & $1$ &  & $\mathcal T_{(0,1)}$ &$=$&
$1\cdot [{\rm t}_2]^+ + 2\cdot [\tfrac{1}{2}]^+ + 1\cdot[0]$
\\[10pt]
$1$ & $3/2$ & long & $\ov{\mathcal S}_{(1)}$ &$=$&
$1\cdot \ov{[\tfrac{3}{2}]}+2\cdot \ov{ [1]} + 4\cdot \ov{[\tfrac{1}{2}]}   +2\cdot \ov{[0]}  +2\cdot \ov{[{\rm t}_2]}$
\\
$1$ & $1$ & long & $\ov{\mathcal V}_{(1)}$ &$=$&
$1\cdot \ov{ [1]} + 2\cdot \ov{[\tfrac{1}{2}]}   +3\cdot \ov{[0]}  $
\vspace*{20pt}
\\
\caption{Supergravity multiplets in $D=6$. The first column shows the amount of
supersymmetry, the second column
indicates the maximal spin of the multiplet, the third column
specifies whether the fields are massless (no indication) or
massive (long or short).
\label{sum_d6}}
\end{longtable}
\end{table}


\clearpage
\subsection{Supergravity in $D=4$}
\label{app_4}

In this section we collect some information about the multiplet structure
in four dimensions. 
The on-shell degrees of freedom of the various fields are summarized as
follows
\eq{
  \renewcommand{\arraystretch}{1.3}
  \begin{array}{l@{\hspace{6pt}}l@{\hspace{20pt}}c@{\hspace{20pt}}c}
  \multicolumn{2}{c}{\mbox{name}} & \mbox{symbol} &\mbox{on-shell d.o.f.}
  \\ \hline\hline
  \mbox{massless}&\mbox{spin~}2&[2]&2_{\rm b}
  \\
  \mbox{massless}&\mbox{spin~}$3/2$&[\tfrac{3}{2}]&2_{\rm f}
  \\
  \mbox{massless}&\mbox{spin~}$1$&[1]&2_{\rm b}
  \\
  \mbox{massless}&\mbox{spin~}$1/2$&[\tfrac{1}{2}]&2_{\rm f}
  \\
  \mbox{massless}&\mbox{spin~}$0$&[0]&1_{\rm b}
  \\[4pt] \hline
  &&&
  \\[-15pt]
  \mbox{massive}&\mbox{spin~}$3/2$&\ov{[\tfrac{3}{2}]}&4_{\rm f}
  \\
  \mbox{massive}&\mbox{spin~}$1$&\ov{[1]}&3_{\rm b}
  \\
  \mbox{massive}&\mbox{spin~}$1/2$&\ov{[\tfrac{1}{2}]}&2_{\rm f}
  \\
  \mbox{massive}&\mbox{spin~}$0$&\ov{[0]}&1_{\rm b}
  \end{array}
}  
In table \ref{multi_01} on pages \pageref{page_01} and \pageref{page_02} the massless and
massive multiplets in four dimensions are summarized.
This data has been taken 
from \cite{Andrianopoli:2002rm} and has been included here for completeness.


\afterpage{
\label{page_01}
\vspace*{10pt}
\setlength{\LTcapwidth}{\linewidth}
\tabcolsep10pt
\renewcommand{\arraystretch}{1.4}
\begin{longtable}{c|c|c|l@{\hspace{2pt}}c@{\hspace{2pt}}l}
$\mathcal N$ & spin & mass & \multicolumn{3}{l}{content}
\\ \hline\hline
$8$ & $2$ &  & $\mathcal G_{(8)}$ &$=$&
$1\cdot [2] + 8\cdot [\tfrac{3}{2}]   +28\cdot [1] + 56\cdot [\tfrac{1}{2}] + 70\cdot [0] $
\\[10pt]  
$6$ & $2$ &  & $\mathcal G_{(6)}$ &$=$&
$1\cdot [2] + 6\cdot [\tfrac{3}{2}]   +16\cdot [1] + 26\cdot [\tfrac{1}{2}] + 30\cdot [0] $
\\
$6$ & $3/2$ &  & $\mathcal S_{(6)}$ &$=$&
$1\cdot [\tfrac{3}{2}]   +6\cdot [1] + 15\cdot [\tfrac{1}{2}] + 20\cdot [0] $
\\
$6$ & $3/2$ & $\tfrac{1}{2}\:$BPS & $\ov{\mathcal S}_{(6)}$ &$=$&
$2\cdot \ov{[\tfrac{3}{2}]}   +12\cdot \ov{[1] }+ 28\cdot \ov{[\tfrac{1}{2}]} + 28\cdot \ov{[0]} $
\\[10pt]  
$5$ & $2$ &  & $\mathcal G_{(5)}$ &$=$&
$1\cdot [2] + 5\cdot [\tfrac{3}{2}]   +10\cdot [1] + 11\cdot [\tfrac{1}{2}] + 10\cdot [0] $
\\
$5$ & $3/2$ &  & $\mathcal S_{(5)}$ &$=$&
$1\cdot [\tfrac{3}{2}]   +6\cdot [1] + 15\cdot [\tfrac{1}{2}] + 20\cdot [0] $
\\
$5$ & $3/2$ & $\tfrac{2}{5}\:$BPS & $\ov{\mathcal S}_{(5)}$ &$=$&
$2\cdot \ov{[\tfrac{3}{2}]}   +12\cdot \ov{[1]} + 28\cdot \ov{[\tfrac{1}{2}]} + 28\cdot \ov{[0]} $
\\[10pt]  
$4$ & $2$ &  & $\mathcal G_{(4)}$ &$=$&
$1\cdot [2] + 4\cdot [\tfrac{3}{2}]   +6\cdot [1] + 4\cdot [\tfrac{1}{2}] + 2\cdot [0] $
\\
$4$ & $3/2$ &  & $\mathcal S_{(4)}$ &$=$&
$1\cdot [\tfrac{3}{2}]   +4\cdot [1] + 7\cdot [\tfrac{1}{2}] + 8\cdot [0] $
\\
$4$ & $3/2$ & $\tfrac{1}{4}\:$BPS & $\ov{\mathcal S}^{1/4}_{(4)}$ &$=$&
$2\cdot \ov{[\tfrac{3}{2}]}   +12\cdot \ov{[1]} + 28\cdot \ov{[\tfrac{1}{2}]} + 28\cdot \ov{[0]} $
\\
$4$ & $3/2$ & $\tfrac{1}{2}\:$BPS & $\ov{\mathcal S}^{1/2}_{(4)}$ &$=$&
$2\cdot \ov{[\tfrac{3}{2}]}   +8\cdot \ov{[1]} + 12\cdot \ov{[\tfrac{1}{2}]} + 8\cdot \ov{[0] }$
\\
$4$ & $1$ &  & $\mathcal V_{(4)}$ &$=$&
$1\cdot [1] + 4\cdot [\tfrac{1}{2}] + 6\cdot [0] $
\\
$4$ & $1$ & $\tfrac{1}{2}\:$BPS & $\ov{\mathcal V}_{(4)}$ &$=$&
$2\cdot \ov{[1]} + 8\cdot \ov{[\tfrac{1}{2}]} + 10\cdot \ov{[0] }$
\\[10pt]  
$3$ & $2$ &  & $\mathcal G_{(3)}$ &$=$&
$1\cdot [2] + 3\cdot [\tfrac{3}{2}]   +3\cdot [1] + 1\cdot [\tfrac{1}{2}] $
\\
$3$ & $3/2$ &  & $\mathcal S_{(3)}$ &$=$&
$1\cdot [\tfrac{3}{2}]   +3\cdot [1] + 3\cdot [\tfrac{1}{2}] + 2\cdot [0] $
\\
$3$ & $3/2$ & long & $\ov{\mathcal S}^{\:l}_{(3)}$ &$=$&
$1\cdot \ov{[\tfrac{3}{2}]}   +6\cdot \ov{[1]} + 14\cdot \ov{[\tfrac{1}{2}] }+ 14\cdot \ov{[0] }$
\\
$3$ & $3/2$ & $\tfrac{1}{3}\:$BPS & $\ov{\mathcal S}^{1/3}_{(3)}$ &$=$&
$2\cdot \ov{[\tfrac{3}{2}]}   +8\cdot\ov{ [1]} + 12\cdot \ov{[\tfrac{1}{2}]} + 8\cdot \ov{[0] }$
\\
$3$ & $1$ &  & $\mathcal V_{(3)}$ &$=$&
$1\cdot [1] + 4\cdot [\tfrac{1}{2}] + 6\cdot [0] $
\\
$3$ & $1$ & $\tfrac{1}{3}\:$BPS & $\ov{\mathcal V}_{(3)}$ &$=$&
$2\cdot \ov{[1] }+ 8\cdot \ov{[\tfrac{1}{2}] }+ 10\cdot\ov{ [0] }$
\\[10pt]  
$2$ & $2$ &  & $\mathcal G_{(2)}$ &$=$&
$1\cdot [2] + 2\cdot [\tfrac{3}{2}]   +1\cdot [1] $
\\
$2$ & $3/2$ &  & $\mathcal S_{(2)}$ &$=$&
$1\cdot [\tfrac{3}{2}]   +2\cdot [1] + 1\cdot [\tfrac{1}{2}] $
\\
$2$ & $3/2$ & long & $\ov{\mathcal S}^{\:l}_{(2)}$ &$=$&
$1\cdot \ov{[\tfrac{3}{2}]}   +4\cdot \ov{[1] }+ 6\cdot\ov{ [\tfrac{1}{2}]} + 4\cdot\ov{ [0] }$
\\
$2$ & $3/2$ & $\tfrac{1}{2}\:$BPS & $\ov{\mathcal S}^{1/2}_{(3)}$ &$=$&
$2\cdot \ov{[\tfrac{3}{2}] }  +4\cdot \ov{[1] }+ 2\cdot \ov{[\tfrac{1}{2}] }$
\\
\pagebreak 
\multicolumn{1}{l}{}\\[20pt]
$2$ & $1$ &  & $\mathcal V_{(2)}$ &$=$&
$1\cdot [1] + 2\cdot [\tfrac{1}{2}] + 2\cdot [0] $
\\
$2$ & $1$ & long & $\ov{\mathcal V}^{\:l}_{(2)}$ &$=$&
$1\cdot \ov{[1]} + 4\cdot \ov{[\tfrac{1}{2}]} + 5\cdot \ov{[0]} $
\\
$2$ & $1$ & $\tfrac{1}{2}\:$BPS & $\ov{\mathcal V}^{1/2}_{(2)}$ &$=$&
$2\cdot \ov{[1] }+ 4\cdot \ov{[\tfrac{1}{2}]} + 2\cdot\ov{[0]}$
\\
$2$ & $1/2$ &  & $\mathcal H_{(2)}$ &$=$&
$2\cdot [\tfrac{1}{2}] + 4\cdot [0] $
\\
$2$ & $1/2$ & $\tfrac{1}{2}\:$BPS & $\ov{\mathcal H}_{(2)}$ &$=$&
$2\cdot\ov{ [\tfrac{1}{2}]} + 4\cdot \ov{[0] }$
\\[10pt]  
$1$ & $2$ &  & $\mathcal G_{(1)}$ &$=$&
$1\cdot [2] + 1\cdot [\tfrac{3}{2}]  $
\\
$1$ & $3/2$ &  & $\mathcal S_{(1)}$ &$=$&
$1\cdot [\tfrac{3}{2}]   +1\cdot [1] $
\\
$1$ & $3/2$ & long & $\ov{\mathcal S}^{\:l}_{(1)}$ &$=$&
$1\cdot \ov{[\tfrac{3}{2}] }  +2\cdot \ov{[1] }+ 1\cdot \ov{[\tfrac{1}{2}] }$
\\
$1$ & $1$ &  & $\mathcal V_{(1)}$ &$=$&
$1\cdot [1] + 1\cdot [\tfrac{1}{2}] $
\\
$1$ & $1$ & long & $\ov{\mathcal V}^{\:l}_{(1)}$ &$=$&
$1\cdot \ov{[1] }+2\cdot \ov{[\tfrac{1}{2}] }+ 1\cdot\ov{ [0] }$
\\
$1$ & $1/2$ &  & $\mathcal C_{(1)}$ &$=$&
$1\cdot [\tfrac{1}{2}] + 2 \cdot [0]$
\\
$1$ & $1/2$ & long  & $\ov{\mathcal C}_{(1)}$ &$=$&
$1\cdot\ov{ [\tfrac{1}{2}] }+ 2 \cdot\ov{ [0]}$ \vspace*{20pt}
\\
\caption{Supergravity multiplets in $D=4$. The first column shows the amount of
supersymmetry, the second column
indicates the maximal spin of the multiplet, the third column
specifies whether the fields are massless (no indication) or
massive (long or shortened).
For more details see \cite{Andrianopoli:2002rm}.
\label{multi_01}}
\end{longtable}
\label{page_02}
\clearpage
}


\clearpage
\bibliography{references}  
\bibliographystyle{utphys}


\end{document}